\documentclass{jfm}
\usepackage{enumitem}
\usepackage{tikz}
\usepackage{graphicx}
\usepackage{epstopdf, epsfig}
\usepackage{amssymb,amsfonts,amsmath}

\definecolor{g-blue}{rgb}{0.83,0.95,1}
\definecolor{g-yellow}{rgb}{1,1,0.7}
\definecolor{g-green}{rgb}{0.9,1,0.9}
\definecolor{green}{rgb}{0,0.6,0}
\definecolor{cyan}{rgb}{0,0.7,0.7} 
\definecolor{grey}{rgb}{0.4 ,0.4 ,0.4 }
\definecolor{brown}{rgb}{0.6 ,0  ,0.8 }

\def\g-blue#1{\textcolor{g-blue}{#1}}

\usepackage[authoryear,square]{natbib}

\shorttitle{Surface gravity wave---interfacial wave interactions}
\shortauthor{ J. Zaleski, P. Zaleski and Y. V. Lvov}

\title{Excitation of interfacial waves via surface---interfacial wave interactions}

\author{Joseph Zaleski  \aff{1},
Philip Zaleski \aff{2},
 \and Yuri V. Lvov \aff{1}
 \corresp{\email{lvovy@rpi.edu}}}

\affiliation{\aff{1}Department of Mathematics, Rensselaer Polytechnic Institute, NY 12180, USA
\aff{2}Department of Mathematics, New Jersey Institute of Technology, Newark, NJ 07102, USA}

\DeclareMathOperator{\sech}{sech}
\begin{document}

\maketitle

\begin{abstract}
  \indent We consider interactions between surface and interfacial
  waves in the two layer system. Our approach is based on the
  Hamiltonian structure of the equations of motion, and includes the
  general procedure for diagonalization of the quadratic part of the
  Hamiltonian. Such diagonalization allows us to derive the
  interaction crossection between surface and interfacial waves and to
  derive the coupled kinetic equations describing spectral energy
  transfers in this system. Our kinetic equation allows resonant and
  near resonant interactions. We find that the energy transfers are
  dominated by the class III resonances of \cite{Alam}. We apply our
  formalism to calculate the rate of growth for interfacial waves for
  different values of the wind velocity. Using our kinetic equation,
  we also consider the energy transfer from the wind generated surface
  waves to interfacial waves for the case when the spectrum of the
  surface waves is given by the JONSWAP spectrum and interfacial waves
  are initially absent.  We find that such energy transfer can occur
  along a timescale of hours; there is a range of wind speeds for the
  most effective energy transfer at approximately the wind speed
  corresponding to white capping of the sea. Furthermore, interfacial
  waves oblique to the direction of the wind are also generated.
\end{abstract}

\begin{keywords}
\end{keywords}

\section{Introduction\label{Intro}}
\subsection{Background}
\indent The term ``ocean waves'' typically evokes images of surface
waves shaking ships during storms in the open ocean, or breaking
rhythmically near the shore. However, much of the ocean wave action
takes place far underneath the surface, and consists of surfaces of
constant density being disturbed and modulated.

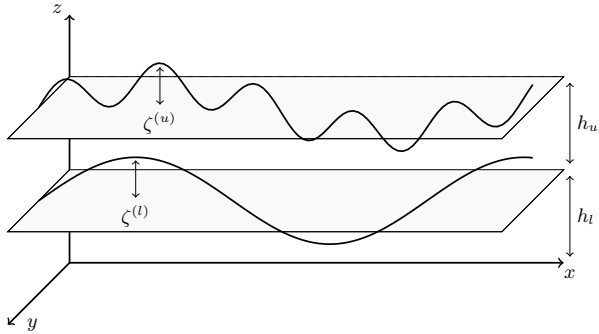
\begin{figure}
\resizebox{.6\linewidth}{!}{
\begin{tikzpicture}
\draw [<->,  thick] (0,4) -- (0,0) -- (8,0);
\draw [->, thick] (0,0) -- (-1,-1);
\node[left] at (0,4.1) {$z$};
\node[below] at (8.1,0) {$x$};
\node[right] at (-0.8,-1) {$y$};
\draw[black, dashed, domain=0:8, samples=100] plot (\x, {3});
\draw[fill=gray!5] (0,3) -- (-1,2) -- (7,2)--(8,3)--(0,3);
\draw[black, thick, domain=0:8, samples=100] plot (\x-0.5, {0.32*sin(4*\x r)+3+0.4*sin(0.85*\x r)-0.5});
\draw[-] (0,3) -- (-1,2) -- (7,2)--(8,3);

\draw[fill=gray!5] (0,1.5) -- (-1,0.5) -- (7,0.5)--(8,1.5)--(0,1.5);
\draw[black,  thick, domain=0:8, samples=100] plot (\x-0.5, {0.7*sin(\x r)+1.5-0.5});
\draw[-] (0,1.5) -- (-1,0.5) -- (7,0.5)--(8,1.5);

\draw [<->] (8.1,0.1) -- (8.1,1.4);
\draw [<->] (8.1,1.6) -- (8.1, 2.9);
\node[right] at (8.1,0.75) {$h_l$};
\node[right] at (8.1,2.25) {$h_u$};
\draw [<->] (5*pi/8-0.5,3.05-0.5) -- (5*pi/8-0.5,3.65-0.5);
\node[below] at (5*pi/8-0.5,3.05-0.5) {$\zeta^{(u)}$};
\draw [<->] (pi/2-0.5,1.55-0.5) -- (pi/2-0.5, 1.85-0.2);
\node[below] at (pi/2-0.5,1.55-0.5) {$\zeta^{(l)}$};
\end{tikzpicture}
}
\caption{Schematic of the surface and interface with respect to mean displacements}
\label{2LayerSystem}
\end{figure}

When wind blows over the ocean, it excites surface waves.  These
surface waves in turn excite the internal waves. Therefore the
coupling between surface and interfacial waves provide a key mechanism
of coupling an atmosphere and the ocean. The simplest conceptual model
describing such interaction is a two layer model figure
(\ref{2LayerSystem}), with the lighter fluid with free surface being
on top of the heavier fluid with the rigid bottom.  This two layer
model has been actively studied in the last few decades from the angle
of weakly nonlinear resonant interactions between surface and
interfacial layers \cite{Ball, Thorpe,GargetttJFM,
  WatsonJFM,OlbersJFM,SegurPOF,DystheDasJFM,Watson1990,Watson1994,Alam,TanakaWakJFM,ConstantinPOF,OlbersJOPO}.
\\
\indent The strength of such nonlinear interactions has been the subject of long debate. Earlier approaches include the calculations of \cite{Thorpe,OlbersJFM}.  Most recently, \cite{OlbersJOPO} found the annual mean energy flux integrated globally over the oceans to be about $10^{-3}$ terawatts. 
\indent \cite{Ball} showed the existence of a closed curve of triad
resonances for two dimensional wave vectors corresponding to
interactions between waves of all possible orientations. He emphasized
the cases in which two counter propagating surface waves drive an
interfacial wave, and two counter propagating interfacial waves drive a
surface wave.  Later, these classes of resonances were referred to as
class I and class II interactions. In class I, two surface waves
counter propagate with roughly equal wavelength, with the interfacial
wave having shorter wavelength. In class II, the interfacial waves
counter propagate with the surface wave having roughly twice the
frequency of the interfacial waves \cite{Alam}.
\\ \indent \cite{Chow} analyzed class I resonances for a two layer
model under the assumptions that the bottom layer is of infinite depth
and the top layer is shallow. These assumptions allowed to formulate
the triad resonance condition in a more general way, namely that the
group speed of a surface wave envelope matches the phase speed of the
interfacial wave.  Using this condition, Chow derived evolution
equations for a surface wave train coupled to interfacial waves. He
found a band of wavenumbers that are unstable, thus facilitating
energy transfer from the surface waves to the interfacial waves.
However, he found that the energy transfer rate was smaller than the
transfer rate for resonant triad interactions.
 
\indent \cite{Watson1990} considered surface-interfacial waves
interaction, taking into account both surface wave dissipation,
and broadening of the three wave  resonances, using WKB theory.
\\ \indent \cite{Alam} considered resonances in the
one-dimensional case with collinear waves.  He discovered what
is now called class III resonances, where the wavelength of the
resonant interfacial wave is much longer than that of the two
co-propagating surface waves, making it physically relevant in
describing the formation of long interfacial waves. Alam showed that
these resonances cause a cascade of resonant and near-resonant
interactions between surface and interfacial waves and thus could be a
viable energy exchange mechanism. He also obtained expressions for the
amplitude growth of an interfacial wave in a system with a large
number of interacting waves.
\\ \indent \cite{TanakaWakJFM} considered the two layer system and
modeled numerically the primitive equations of motion in 2+1
dimensions (horizontal and vertical direction and time) for the case
of a surface wave spectrum based on the Pierson---Moskowitz spectrum
\cite{Pierson1964}. \cite{TanakaWakJFM} showed that the initially
still interface experiences excitation with a flux of energy towards
smaller wave numbers.  For the case of large difference in density
between layers, they noticed that the shape of the surface spectrum
changes significantly. They noted that this cannot be explained by
resonant wave interaction theory because resonant wave interaction
theory predicts the existence of the critical surface wave number,
below which there could not be any interactions. Consequently, there
is a need of a theory not limited to only resonant, but also near
resonant interactions.
\\ \indent \cite{OlbersJOPO} used an analytical framework that
directly derives the flux of energy radiating downward from the mixed
layer base with a goal of providing a global map of the energy
transfer to the interfacial wave field.  They concluded that
spontaneous wave generation, where two surface waves create an
interfacial wave, becomes dominant over modulational interactions
where a preexisting interfacial wave is modulated by a surface wave
for wind speeds above 10-15 m/s.\\
\subsection{Overview of the paper}
\indent In this paper we derive from first principle wave turbulence
theory for wave-wave interactions in the two-layer model. Our theory
is based on the recently derived Hamiltonian structure for this system
~\cite{Choi2019}. We derive the kinetic equations describing weakly
nonlinear energy transfers between waves. The theory includes both
resonant and near-resonant wave-wave interactions, and allows to
quantitatively describe coupling between the atmosphere and the ocean.
\\
  \indent The paper is written as follows: in section \ref{GovEq} we
discuss the governing equations of motion and Hamiltonian structure
derived by \cite{Choi2019} for the case of 3+1 dimensions. Notably,
the Hamiltonian is expressed explicitly in terms of the interface
variables, forming the base needed for our analysis.
\\ \indent In section \ref{CanTrans} we derive a canonical
transformation to diagonalize the quadratic part of the Hamiltonian to
obtain the normal modes. Such a diagonalization reduces the system to
an ensemble of waves which are free to the leading order, thus making
it amendable to Wave Turbulence theory as described in
\cite{ZLF,NazBook}.
\\ \indent In section \ref{StatApproach} we apply Wave Turbulence
theory to obtain the system of kinetic equations governing the time
evolution of the wave action spectrum of waves (``number of
waves"). Furthermore, we calculate the exact matrix elements
(interaction crossections) governing such interactions. Our
calculations are valid both on and near the resonant--manifold.
\\ \indent In section \ref{EnergyTransferJONSWAPP} we test our theory
by considering the model problem with the surface described by a
single plain wave with frequency being the peak frequency of the
JONSWAP spectrum \cite{JONSWAP}.  We obtain the Boltzmann rates and
corresponding timescales of amplitude growth for excited interfacial
waves.  The frequencies of the excited interfacial waves are well
within the experimental measurements given in buoyancy profiles of the
ocean. Furthermore, we generalize the results in \cite{Alam} by
considering the general case where all wave vectors are two
dimensional and not necessarily collinear. Notably, for conditions of
long surface swell waves, the dominant interactions occur between
surface and interfacial waves which are oblique, a case also noted by
\cite{HaneyJFM}.  Inspired by \cite{TanakaWakJFM}, we also simulate
the evolving spectra for the case when the surface is the 1-D JONSWAP
spectrum and the interface is initially at rest.
\\
\indent In section \ref{Discuss} we conclude by summarizing our results and discussing future work.  

\section{Governing equations}\label{GovEq}
\subsection{Equations in physical space} 
We use a Cartesian coordinate system $(\boldsymbol{x},z)$, with the
$xy$-plane being the mean free surface and the $z$-axis being directed
upward.  We consider a two-layer model with the free surface on top
and the interface between the layers.  We denote the respective depths
of the upper and lower layers by $h_u, h_l$ and their densities by
$\rho_u$, $\rho_l$, with the difference in density between layers to
be $\Delta \rho= \rho_l - \rho_u$, where subscript ``u'' refers to the
upper and subscript ``l'' to lower layers.  We assume the fluid is
homogeneous, incompressible, immersible, inviscid, and irrotational in
both layers.

To derive the closed set of coupled equations for the surface velocity
potential $\Psi^{(u)}(x,y,t)$ and displacement $\zeta^{(u)}(x,y,t)$
and interfacial velocity potential $\Psi^{(l)}(x,y,t)$ and
displacement $\zeta^{(l)}(x,y,t)$ we start from the Euler equations,
incompressibility condition, and kinematic boundary conditions for
velocity and pressure continuity along the surface/interface. We then
introduce a nonlinearity parameter $\epsilon$, the slope of the waves,
and make a formal assumption that $\epsilon\ll 1$. This allows us to
iterate the resulting equations for a solution representing a
wavetrain with wave number $k$.  This procedure was recently executed
in \cite{Choi2019}, and leads to the following system of equations,
truncated at the second order of nonlinearity parameter,
\begin{subequations}
\label{eqm}
\begin{align}
&\dot{\zeta}^{(u)}={{\gamma_{11}}} \Psi^{(u)}
+\gamma_{12} \Psi^{(l)}-\rho_u \gamma_{11} \big[ \zeta^{(u)} (\gamma_{11} \Psi^{(u)} + \gamma_{12} \Psi^{(l)} \big]
-\Delta \rho \gamma_{21}\big[\zeta^{(l)} ( \gamma_{21} \Psi^{(u)} + \gamma_{22} \Psi^{(l)})\big]
\nonumber
\\
&- \nabla \cdot (\zeta^{(u)} \nabla \Psi^{(u)})/\rho_u
+ \Delta \rho \:(\rho_l/\rho_u)\gamma_{31} \nabla \cdot (\zeta^{(l)} \gamma_{31} \nabla \Psi^{(u)}) -\rho_l {{\gamma_{31}}}\nabla \cdot (\zeta^{(l)} \gamma_{33} \nabla \Psi^{(l)}), 
\\
\nonumber 
\\
&\dot{\zeta}^{(l)}={{\gamma_{21}}} \Psi^{(u)}
+\gamma_{22} \Psi^{(l)}-\rho_u \gamma_{12} \big[ \zeta^{(u)} {{(\gamma_{11} \Psi^{(u)} + \gamma_{12} \Psi^{(l)} )}}\big]
-\Delta \rho \gamma_{22}\big[\zeta^{(l)} ( \gamma_{21} \Psi^{(u)} + \gamma_{22} \Psi^{(l)})\big]
\nonumber
\\
&- \rho_l \gamma_{33} \nabla \cdot (\zeta^{(l)} \gamma_{31} \nabla \Psi^{(u)} ) 
-\rho_l J \nabla \cdot (\zeta^{(l)} J \nabla \Psi^{(l)}) 
+ \rho_u \gamma_{32} \nabla \cdot (\zeta^{(l)} \gamma_{32} \nabla \Psi^{(l)}), 
\\
\nonumber  
\\
&\dot{\Psi}^{(u)} = -\rho_u g \zeta^{(u)} +\frac{1}{2}  \rho_u (\gamma_{11} \Psi^{(u)} + \gamma_{12} \Psi^{(l)})^2 
-\frac{1}{2}( \nabla \Psi^{(u)})\cdot( \nabla \Psi^{(u)})/ \rho_u , 
\\
&\dot{\Psi}^{(l)} = -\Delta \rho g \zeta^{(l)} +\frac{1}{2} \Delta \rho (\gamma_{21} \Psi^{(u)} + \gamma_{22} \Psi^{(l)})^2 
+\frac{1}{2} \Delta \rho (\rho_l / \rho_u) \:(\gamma_{31} \nabla \Psi^{(u)})\cdot(\gamma_{31} \nabla \Psi^{(u)})
\nonumber 
\\
&-\frac{1}{2}\rho_l (J \nabla \Psi^{(l)})\cdot (J \nabla \Psi^{(l)})
+\frac{1}{2}\rho_u (\gamma_{32} \nabla \Psi^{(l)})\cdot(\gamma_{32} \nabla \Psi^{(l)})
-\rho_l (\gamma_{31} \nabla \Psi^{(u)}) \cdot (\gamma_{33} \nabla \Psi^{(l)}), 
\end{align}
\end{subequations}
with the nonlocal linear operators $\gamma_{ij}$ and $J$, whose
Fourier kernels are given in Appendix (\ref{gamJ}).
\subsection{Hamiltonian}
 We use the Fourier transformations of the interface variables for the two layer system depicted in figure (\ref{2LayerSystem}) 
 \begin{equation}
	\zeta^{(j)}(\boldsymbol{x},t)=\int \hat{\zeta}^{(j)}(\boldsymbol{k},t)e^{-i\boldsymbol{k}\cdot \boldsymbol{x}} \text{d}\boldsymbol{k},\:\:\Psi^{(j)}(\boldsymbol{x},t)=\int \hat{\Psi}^{(j)}(\boldsymbol{k},t)e^{-i\boldsymbol{k}\cdot \boldsymbol{x}} \text{d}\boldsymbol{k}\:\text{for}\:j\in\{u,l\}. 
\end{equation}
The equations of motion \eqref{eqm} can then be represented by
canonically conjugated Hamilton's equations for the Hamiltonian $H$,
given by \cite{Choi2019}
\begin{align*}
    \frac{\partial \hat{\zeta}^{(j)}}{\partial t}=\frac{\delta H}{{\delta \hat{\Psi}^{(j)}}^*}, \ \ 
    \frac{\partial \hat{\Psi}^{(j)}}{\partial t}=-\frac{\delta H}{{\delta \hat{\zeta}^{(j)}}^*},\ \ \  \: j \in \{u,l\}. 
\end{align*}

This is a generalization for two layers of the Hamiltonian formulation
described in \cite{zakharov68} for surface waves.  Here the
Hamiltonian, $H$, is a sum of a quadratic Hamiltonian, describing
linear noninteracting waves, and a cubic Hamiltonian, describing
wave--wave interactions, $H=H_2+H_3$, where
\begin{align}
&H_2=\frac{1}{2}\iint \big[ h^{(1a)}_1 \hat{\zeta}^{(u)}_1 \hat{\zeta}^{(u)}_2+h^{(2a)}_1 \hat{\Psi}^{(u)}_1 \hat{\Psi}^{(u)}_2 
\\
\nonumber
&+ h^{(3a)}_1 \hat{\zeta}^{(l)}_1 \hat{\zeta}^{(l)}_2 + h^{(4a)}_1 \hat{\Psi}^{(l)}_1 \hat{\Psi}^{(l)}_2 + h^{(5a)}_{1,2} \hat{\Psi}^{(u)}_1 \hat{\Psi}^{(l)}_2\big]\delta(\boldsymbol{k}_1+\boldsymbol{k}_2) \text{d}\boldsymbol{k_1}\text{d}\boldsymbol{k_2},
\nonumber
\\
  \label{H2}\\
  &H_3=\iiint \big[h^{(1)}_{123}\hat{\Psi}^{(u)}_1 \hat{\Psi}^{(u)}_2 \hat{\zeta}^{(u)}_3 + h^{(2)}_{123}\hat{\Psi}^{(u)}_1 \hat{\Psi}^{(l)}_2 \hat{\zeta}^{(u)}_3 + h^{(3)}_{123}\hat{\Psi}^{(l)}_1 \hat{\Psi}^{(l)}_2 \hat{\zeta}^{(u)}_3
    \nonumber\\
  & +h^{(4)}_{123}\hat{\Psi}^{(u)}_1 \hat{\Psi}^{(u)}_2 \hat{\zeta}^{(l)}_3 + h^{(5)}_{123}\hat{\Psi}^{(u)}_1 \hat{\Psi}^{(l)}_2 {{\hat{\zeta}^{(l)}_3}} + h^{(6)}_{123} \hat{\Psi}^{(l)}_1 \hat{\Psi}^{(l)}_2 \hat{\zeta}^{(l)}_3 \big] \delta(\boldsymbol{k}_1+\boldsymbol{k}_2+\boldsymbol{k}_3) \text{d}\boldsymbol{k_1}\text{d}\boldsymbol{k_2}\text{d}\boldsymbol{k_3}, 
    \nonumber\\ \label{H3}
\end{align}
where the coupling coefficients $h^i_j$  are given in Appendix
\eqref{MatrixElementFS}. Here $k=|\boldsymbol{k}|$ denotes the
wavenumber and we use the notation that subscripts represent vector
arguments, i.e.  $h_{ijl}\equiv h(\boldsymbol{k}_i,\boldsymbol{k}_j,
\boldsymbol{k}_l)$.

This Hamiltonian is expressed explicitly in terms of the variables at
the surfaces of the fluids, and is a significant step forward over the
Hamiltonian structure of the two layer system derived in
\cite{Ambrosi2000}, where the implicit form of the Hamiltonian was
obtained.

The Hamiltonian provides the firm theoretical foundation to develop
the theory of weak nonlinear interactions of surface and interfacial
waves. However to describe the time evolution of the spectral energy
density of the waves, the quadratic part of the Hamiltonian of the
system (\ref{H2}) needs to be diagonalized, so that the linear part of
equations of motion corresponds to distinct noninteracting linear
waves. In other words, we need to calculate the normal modes of the
system. This task is done in the next section.

\section{Canonical transformation to normal modes}\label{CanTrans}
In this paper we use the wave turbulence formalism
~\cite{Newell,N1,Ben,K,ZLF,NazBook} to derive the coupled set of
kinetic equations, describing the spectral energy transfers in the
coupled system of surface and interfacial waves. First we need to
diagonalize the Hamiltonian equations of motion in wave action
variables so that waves are free to the leading order.  This is done
via two canonical transformations; the first being a transformation
from interface variables to complex action density variables done in
subsection (\ref{FirstCan}), the second being a transformation to
diagonalize the quadratic Hamiltonian, giving waves which are free to
the leading order, performed in Section (\ref{TransformedHam}). The
final form of the Hamiltonian in terms of the normal modes is also
derived
in subsection (\ref{TransformedHam}).\\

\subsection{Transformation to complex field variable\label{FirstCan}}
We start from the surface variables for the Fourier image of
displacement of the upper and lower layers
$\hat{\zeta}^{(i)}_{\boldsymbol{k}}$ and the Fourier image of the
velocity potential on upper and lower surfaces
$\hat{\Psi}^{(i)}_{\boldsymbol{k}}$ , where $(i)$ denotes the layer,
with $(u)$ being the upper, and $(l)$ being the lower layers. We then
perform a canonical transformation to complex action variables
describing the complex amplitude of wave with wavenumber
$\boldsymbol{k}$,
\begin{align*}
  &\hat{\zeta}^{(u)}_{\boldsymbol{k}}
  =\Bigg(\frac{{h^{(2a)}_{\boldsymbol{k}}}}{4 h^{(1a)}_{\boldsymbol{k}}}\Bigg)^{1/4} (a^{(u)}_{\boldsymbol{k}}+a^{(u)*}_{\boldsymbol{-k}}),\:\:
\hat{\Psi}^{(u)}_{\boldsymbol{k}}=i\Bigg(\frac{{h^{(1a)}_{\boldsymbol{k}}}}{4 h^{(2a)}_{\boldsymbol{k}}}\Bigg)^{1/4}(a^{(u)}_{\boldsymbol{k}}-a^{(u)*}_{\boldsymbol{-k}}),
\\
&\hat{\zeta}^{(l)}_{\boldsymbol{k}}=\Bigg(\frac{{h^{(4a)}_{\boldsymbol{k}}}}{4 h^{(3a)}_{\boldsymbol{k}}}\Bigg)^{1/4} (a^{(l)}_{\boldsymbol{k}}+a^{(l)*}_{\boldsymbol{-k}}),\:\:
\hat{\Psi}^{(l)}_{\boldsymbol{k}}=i\Bigg(\frac{{h^{(3a)}_{\boldsymbol{k}}}}{4 h^{(4a)}_{\boldsymbol{k}}}\Bigg)^{1/4}(a^{(l)}_{\boldsymbol{k}}-a^{(l)*}_{\boldsymbol{-k}}).
\end{align*}
\indent In these variables the Hamiltonian takes the form 
\begin{align}
\label{HTran1}
&H_2=\int\big[F^{(1)}_{\boldsymbol{k}}|a^{(U)}_{\boldsymbol{k}}|^2+F^{(2)}_{\boldsymbol{k}}|a^{(L)}_{\boldsymbol{k}}|^2+F^{(3)}_{\boldsymbol{k}}[(a^{(U)}_{\boldsymbol{k}}a^{(L)}_{\boldsymbol{-k}}-a^{(U)}_{\boldsymbol{k}}a^{(L)*}_{\boldsymbol{k}})+c.c.]\big] \text{d}{\boldsymbol{k}}, 
\end{align}
\begin{align}
&H_3=\sum_{S_1, S_2, S_3\in\{U,L\}}\iiint \text{d}\boldsymbol{k_1} \text{d}\boldsymbol{k_2}\text{d}\boldsymbol{k_3} \nonumber
\\
&\times \big[(V^{(S_1 S_2 S_3)}_{123}a^{(S_1)*}_1{a^{(S_2)*}_2}a^{(S_3)}_3\delta_{1+2-3}+
G^{(S_1 S_2 S_3)}_{123}a^{(S_1)}_1{a^{(S_2)}_2}a^{(S_3)}_3\delta_{1+2+3}) 
+ c.c.\big].
\end{align}
This is the standard form of the Wave Turbulence Hamiltonian of the
spatially homogeneous nonlinear system with two types of waves and
with the quadratic nonlinearity. The corresponding canonical equations
of motion assume standard canonical form~\cite{ZLF},
\begin{align}
&i \dot{a_{\boldsymbol{k}}}^{(U)} = \frac{\delta H}{\delta a^{(U)*}_{\boldsymbol{k}}}, \ \ 
 i \dot{a_{\boldsymbol{k}}}^{(L)}=\frac{\delta H}{\delta a^{(L)*}_{\boldsymbol{k}}}.
\end{align}

Here the coefficient functions are given by
\begin{align*}
F^{(1)}_{\boldsymbol{k}}=\sqrt{h^{(1a)}_{\boldsymbol{k}} h^{(2a)}_{\boldsymbol{k}}},
F^{(2)}_{\boldsymbol{k}}=\sqrt{h^{(3a)}_{\boldsymbol{k}} h^{(4a)}_{\boldsymbol{k}}},
F^{(3)}_{\boldsymbol{k}} =-\frac{h^{(5)}_{{\boldsymbol{k}},-{\boldsymbol{k}}}}{4}\Big[\frac{h^{(1a)}_{\boldsymbol{k}} h^{(3a)}_{\boldsymbol{k}}}{h^{(2a)}_{\boldsymbol{k}} h^{(4a)}_{\boldsymbol{k}}}\Big]^{1/4}, 
\end{align*}
with the matrix elements $V^{(S_1 S_2 S_3)}_{123}$ and $G^{(S_1 S_2 S_3)}_{123}$ given in appendix \ref{ME}.

\subsection{Hamiltonian in terms of normal modes amplitudes\label{TransformedHam}}

We now need to diagonalize quadratic part of the resulting
Hamiltonian.  We perform this task by finding a canonical
transformation that would decouple linear waves of the Hamiltonian
\eqref{HTran1}. In other words, we are seeking a canonical
transformation to remove the term
$F^{(3)}_{\boldsymbol{k}}[(a^{(U)}_{\boldsymbol{k}}a^{(L)}_{\boldsymbol{-k}}-a^{(U)}_{\boldsymbol{k}}a^{(L)*}_{\boldsymbol{k}})+c.c.]$
from the Hamiltonian \eqref{HTran1}.  The transformation is given by
the equation \eqref{finaltrans} of the Appendix (\ref{ME}).  As a
result we obtain the normal modes of the system while maintaining the
canonical structure of the equations of motion.  Finding such a
transformation to determine the normal modes of the system appears to
be nontrivial task, since we needed to solve over determined system of
nonlinear algebraic equations. Details of this procedure are explained
in Appendix (\ref{ME}). Applying the transformation \eqref{finaltrans}
to equation \eqref{H2}, the quadratic part of the Hamiltonian assumes
the desired form
\begin{align*}
H_2=\int [\tilde{\omega}^{(S)}_{\boldsymbol{k}}|c^{(S)}_{\boldsymbol{k}}|^2 + \tilde{\omega}^{(I)}_{\boldsymbol{k}} |c^{(I)}_{\boldsymbol{k}}|^2] \text{d}{\boldsymbol{k}}, 
\end{align*}
where the superscripts  $S$ and $I$ correspond to the respective surface or
interfacial normal modes.

The linear dispersion relationships of the surface and interfacial normal
modes  are given by
\begin{subequations} 
\begin{align}
&\tilde{\omega}^{(I)}_{\boldsymbol{k}}=\alpha_{\boldsymbol{k}}\cosh 2\phi_{\boldsymbol{k}}+2\gamma_{\boldsymbol{k}} \sinh 2 \phi_{\boldsymbol{k}},
\\
&\tilde{\omega}^{(S)}_{\boldsymbol{k}}=\beta_{\boldsymbol{k}}\cosh2\psi_{\boldsymbol{k}}+2\zeta_{\boldsymbol{k}}\sinh 2\psi_{\boldsymbol{k}}, 
\label{dispersion} 
\end{align}
\end{subequations} 
which can be shown to be equivalent to those in \cite{Choi2019} and
\cite{Alam}. 
    The transformation
    \eqref{finaltrans} also alters the higher order terms of the Hamiltonian
    due to three-wave interactions
\begin{align*}
H_3=&\sum_{S_1, S_2, S_3\in\{S,I\}}\iiint \text{d}\boldsymbol{k_1}\text{d}\boldsymbol{k_2}\text{d}\boldsymbol{k_3}
\\
&\times \big[(J^{(S_1 S_2 S_3)}_{123}c^{(S_1)*}_1{c^{(S_2)*}_2}c^{(S_3)}_3\delta_{1+2-3}+
L^{(S_1 S_2 S_3)}_{123}c^{(S_1)}_1{c^{(S_2)}_2}c^{(S_3)}_3\delta_{1+2+3})
+ c.c.\big].
\end{align*}
Here $J^{(S_1 S_2 S_3)}_{123}$ and $L^{(ijk)}_{123}$ are the
interaction matrix elements, also called scattering
crossections. These matrix elements describe the strength of the
nonlinear coupling between wave numbers of $\boldsymbol{k}_1$,
$\boldsymbol{k}_2$ and $\boldsymbol{k}_3$ of the normal modes of types
$S_1$, $S_2$ and $S_3$. Calculation of these matrix elements is a
tedious but straightforward task completed in Appendix
\eqref{METrans}. {{An alternative, but equivalent, formulation is
    described in detail for both the resonant and near--resonant cases
    in \cite{Choi2019}.}} Knowledge of these matrix elements and the
linear dispersion relations allows us to use the wave turbulence
formalism to derive the kinetic equations describing the time
evolution of the spectral energy density of interacting waves. This is
done in the next Section.

\section{Statistical approach via wave turbulence theory}\label{StatApproach}
In wave turbulence the system is represented as a superposition of
large $N$ waves with the complex amplitudes,
$c^{(S)}_{\boldsymbol{k}}(t)$, $c^{(I)}_{\boldsymbol{k}}(t)$,
interacting with each other. In its essence, the classical wave
turbulence theory is a perturbation expansion of the complex wave
amplitudes in terms of the nonlinearity, yielding, at the leading
order, linear waves, with amplitudes slowly modulated at higher orders
by resonant nonlinear interaction.  This modulation leads to a
resonant or near-resonant redistribution of the spectral energy
density among length-scales, and is described by a system of kinetic
equations, the time evolution equations for the wave spectra of
surface and interfacial waves, respectively
\begin{subequations}
\begin{align}
    &n^{(S)}({\boldsymbol{k}},t)\delta({\boldsymbol{k}}-{{\boldsymbol{k}}}')=\langle c^{(S)}_{\boldsymbol{k}} {c^{(S)}_{{\boldsymbol{k}}'}}^*\rangle,
    \\
    &n^{(I)}(\boldsymbol{k},t)\delta({\boldsymbol{k}}-{{\boldsymbol{k}}}')=\langle c^{(I)}_{\boldsymbol{k}} {c^{(I)}_{{\boldsymbol{k}}'}}^*\rangle,
\end{align}
\end{subequations} 
where $\langle...\rangle$ denotes an ensemble average over all
possible realizations of the systems.

Wave turbulence theory has led to spectacular success in predicting
spectral energy densities in the ocean, atmosphere, and plasma, see 
\cite{ZLF,NazBook} for review.

\subsection{Kinetic equations} 
The kinetic equation is the classical analog of the Boltzmann
collision integral. The basic ideas for writing down the kinetic
equation to describe how weakly interacting waves share their energies
go back to Peierls. The modern theory has its origin in the works of
Hasselmann, Benney and Saffmann, Kadomtsev, Zakharov, and Benney and
Newell.  \\ \indent There are many ways to derive the kinetic equation
which are well understood and well studied,
\cite{Ben,ZLF,X1,X2,X3,LN,NazBook}. Here we use the slightly more
general approach for the derivation of the kinetic equations, which
allows not only for resonant, but also for near resonant interactions,
as it was done in \cite{LLNZ,LvovNearRes}. We generalize
\cite{LLNZ,LvovNearRes} for the case of a system of two types of
interacting waves, namely surface and interfacial waves. The resulting
system of kinetic equations is
\begin{align}
  \dot{n}^{(S_0)}({\boldsymbol{k}},t)&=&\sum_{S_1, S_2 \in \{S,I\}}
  \iint \text{d}\boldsymbol{k_1}\text{d}\boldsymbol{k_2}\times
  \Big(|J^{(S_0 S_1 S_2)}_{012}|^2 f^{(S_0 S_1 S_2)}_{012} \delta(\boldsymbol{k}-\boldsymbol{k}_1-\boldsymbol{k}_2) \mathcal{L}^{(S_0 S_1 S_2)}_{{\bf k},{\bf k_1},{\bf k_2}} 
\nonumber
\\
&& \ \ \ \ \ \  \ \ \ \ \ 
- 2 |J^{(S_1 S_0 S_2)}_{102}|^2 f^{(S_1 S_0 S_2)}_{102}\delta(\boldsymbol{k}_1-\boldsymbol{k}-\boldsymbol{k}_2) \mathcal{L}^{(S_1 S_0 S_2)}_{{\bf k_1},{\bf k}, {\bf k_2}} \Big), \nonumber\\ &&S_0\in \{S,I\}, \nonumber\\
\label{KE} 
\end{align}
where $f^{(S_1 S_2 S_3)}$ is the kernel  
three-wave kinetic equation kernel for two types of
waves
\begin{align*}
    f^{(S_1 S_2 S_3)}_{123}= n^{(S_1)}_1 n^{(S_2)}_2 n^{(S_3)}_3\big(\frac{1} {n^{(S_1)}_1} -\frac{1} {n^{(S_2)}_2}-\frac{1} {n^{(S_3)}_3}\big). 
\end{align*}
The frequency conserving Dirac Delta function is replaced by the its
broadened version, the Lorenzian ${\cal L}$  
\begin{align}
  &{\cal{L}}^{(S_1 S_2 S_3)}_{{\bf k},{\bf k_1},{\bf k_2}}=
  \frac{\Gamma^{(S_1 S_2 S_3)}_{\boldsymbol{k}_1, \boldsymbol{k}_2, \boldsymbol{k}_3}}
       {\left(\omega_{\bf k}^{(S_1)}-\omega_{{\bf k_2}}^{(S_2)}
         - \omega_{\bf k_2}^{(S_2)}\right)^2+
         \left(\Gamma^{(S_1 S_2 S_3)}_{\boldsymbol{k}, \boldsymbol{k_1}, \boldsymbol{k_2}}\right)^2}, 
    \nonumber\\
    &\Gamma^{(S_1 S_2 S_3)}_{\boldsymbol{k}_1, \boldsymbol{k}_2, \boldsymbol{k}_3}=
     \gamma^{(S_i)}_1+ \gamma^{(S_i)}_2+ \gamma^{(S_i)}_3,
    \nonumber\\
    &\gamma^{(S_0)}_{\bf k}= \sum_{S_1, S_2 \in \{S,I\}}\iint \text{d}\boldsymbol{k_1}\text{d}\boldsymbol{k_2}
\Big(|J^{(S_0 S_1 S_2)}_{012}|^2 (n^{(S_1)}_1 + n^{(S_2)}_2)\delta(\boldsymbol{k}-\boldsymbol{k}_1-\boldsymbol{k}_2)
    \mathcal{L}^{(S_0 S_1 S_2)}_{{\bf k},{\bf k_1},{\bf k_2}} 
\nonumber 
\\
&\hskip 4.5cm- 2 |J^{(S_1 S_0 S_2)}_{102}|^2 (n^{(S_1)}_1-n^{(S_2)}_2)\delta(\boldsymbol{k}_1-\boldsymbol{k}-\boldsymbol{k}_2) \mathcal{L}^{(S_1 S_0 S_2)}_
  {{\bf k_1},{\bf k},{\bf k_2}}  \Big),
\nonumber \\  &\hskip 10cm S_0\in \{S,I\}.\nonumber \\
\label{gamma}
\end{align}
Here $\Gamma^{(j)}_{\boldsymbol{k}_1, \boldsymbol{k}_2,
\boldsymbol{k}_3}$ is the total broadening of a given resonance
between wavenumbers $\boldsymbol{k}_1, \boldsymbol{k}_2,
\boldsymbol{k}_3$ , and $\gamma^{(S_i)}_i$ is the Boltzmann rate for
wavevector $(\boldsymbol{k}_i, S_i)$.

\indent The principle new feature of this system of kinetic equations
is that instead of the resonant interactions taking place along
Dirac-delta functions, the near-resonances appear acting along the
broadened resonant manifold that includes not only resonant, but also
near resonant interactions, as it was done in
\cite{LLNZ,LvovNearRes}.

The interpretation of this formula is the following: nonlinear
wave-wave interactions lead to the change of wave amplitude, which in
turn makes the lifetime of the waves to be finite. Consequently,
interactions can be near-resonant.  A self-consistent evaluation of
$\gamma_{\bf{k}} $ requires the iterative solution of (\ref{KE}) and
(\ref{gamma}) over the entire field. Indeed, one can see from
(\ref{gamma}) that the width of the resonance depends on the lifetime
of an individual wave, which in turn depends on the resonance width
over which wave interactions occur.

We define our characteristic time for interfacial wave growth to be
$\tau^{(S_i)}_{i}=\frac{-1}{\gamma^{(S_i)}_i}$, i.e. the
$\text{e}$-scaling rate of the action density variable $n^{(S_i)}_i$.
Together \eqref{KE}---\eqref{gamma} form a closed set of equations
which can be iteratively solved to obtain the time evolution of the
energy spectrum of surface and interfacial waves.
\subsection{Three wave resonances}

\begin{figure}
	\begin{flushleft}
          \includegraphics[width=2.5in]{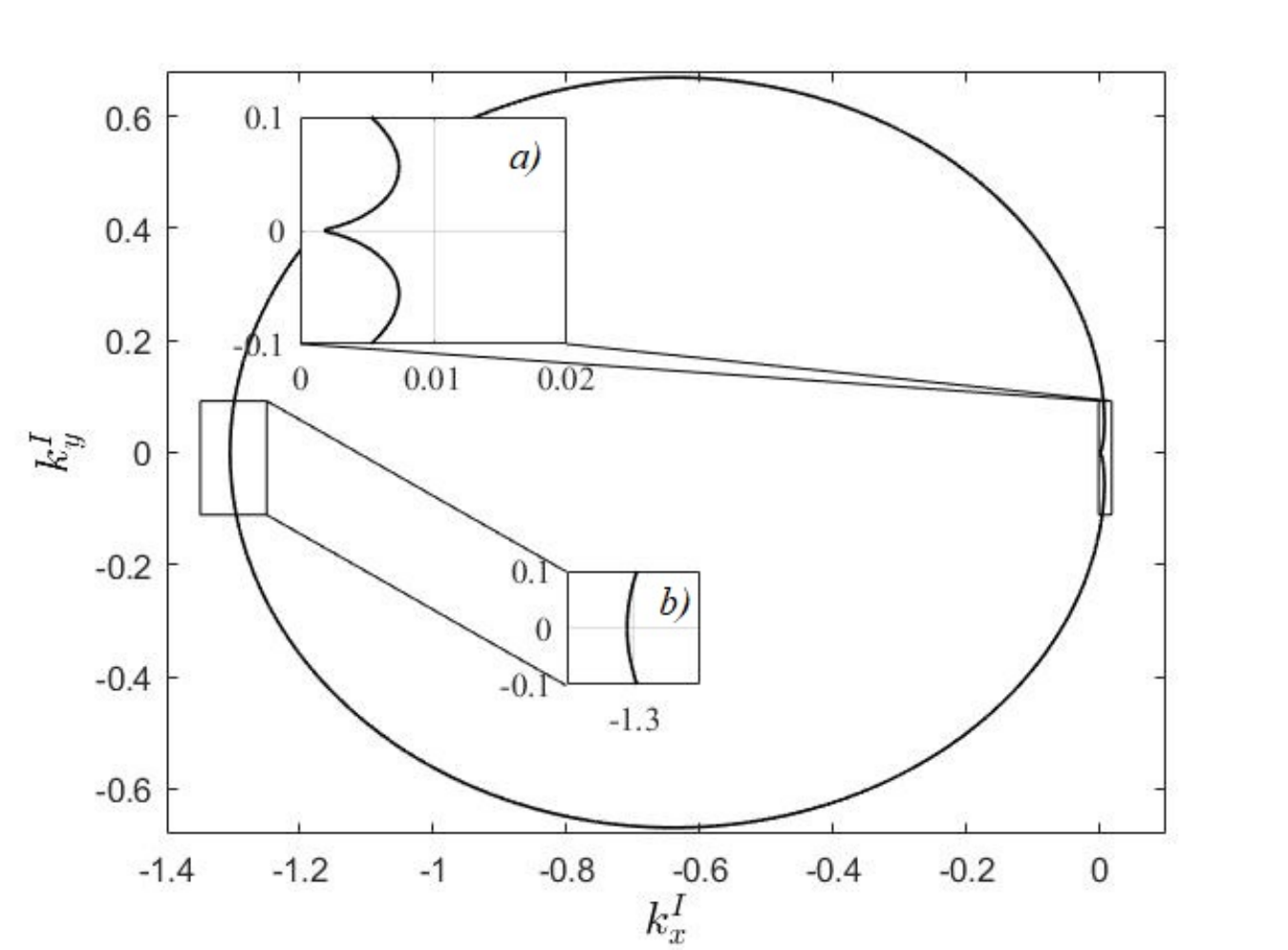}
          \caption{The resonant set of interfacial wave numbers for
            fixed surface wave number $\kappa=\frac{2 \pi}{10}$
            $\text{m}^{-1}$.  (a) class III and surrounding
            resonances, (b) class I and surrounding resonances}
    \label{res}
	\end{flushleft}
      \end{figure}

\begin{figure}
\begin{tikzpicture}
\draw[thick] (0,0) circle (2cm);
\draw[->,  thick] (2,0) -- (8,0);
\draw[->,  thick] (2,0) -- (0,2);
\draw[->,  thick] (2,0) -- (10,-2);
\draw[thick, dashed] (0,2) -- (8,0);
\draw[thick, dashed] (10,-2) -- (8,0);

\draw[->,  thin] (2,0) -- (0,-2);

\draw[->,  thin] (2,0) -- (10,2);

\foreach \Point in {(2,0),(8,0),(0,2),(10,-2),(0,-2),(10,2)}{
    \node at \Point {\textbullet};
}

\node[below] at (2.4,-0.1) {$(0,0)$};

\node[below] at (6.5,0) {$\boldsymbol{\kappa^{(S)}}$};
\node[below] at (0.5,1.3) {${\boldsymbol{k}}^{(I)}_2$};
\node[below] at (6,-1.1) {${\boldsymbol{\kappa}^{(S)}}-{\boldsymbol{k}}^{(I)}_2 $};
\end{tikzpicture}
\caption{Schematic construction of three-wave resonances as determined
  by a fixed surface wave, $\boldsymbol{\kappa}$ and the corresponding
  set of resonant interfacial waves, as done in \cite{Ball}.  Two of
  such triads are depicted.}
\label{resonanceSchematic}
\end{figure}
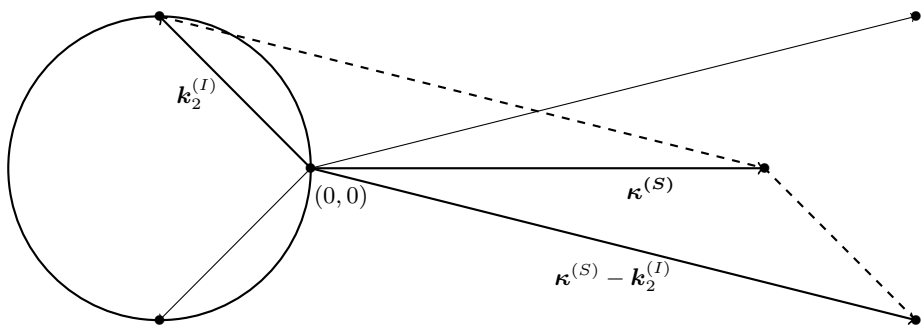
Wave turbulence theory considers the resonant wave-wave interactions.
The rationale for this is that out of all possible interactions of
three wave numbers it is only resonant and near-resonant interactions
that lead to the effective irreversible energy exchange between wave
numbers. Consequently, we seek wave numbers that satisfy the following
resonances of the form
\begin{align}
&{\boldsymbol{k}}_1\pm{\boldsymbol{k}}_2\pm{\boldsymbol{k}}_3=0,
\label{resk} 
\\
&{\omega}^{(S_1)}_1\pm{\omega}^{(S_2)}_2\pm{\omega}^{(S_3)}_3=0, \:\:\: S_1, S_2, S_3 \in \{S,I\}.  
\label{resfreq}
\end{align}
\\
\indent In figure \eqref{res} we plot an example of the 2-D resonant
set as described in \cite{Ball, Thorpe}.  We first fix the surface wave
number $\boldsymbol{k}^S_1$ to be a fixed, given number. We
arbitrarily choose $\boldsymbol{k}^S_1= 2\pi/10$, and calculate wave
numbers $\boldsymbol{k}^I_2$ and $\boldsymbol{k}^S_3$ so that the
conditions $\boldsymbol{k}^S_1=\boldsymbol{k}^I_2+\boldsymbol{k}^S_3$
and ${\omega}^{(S)}_1={\omega}^{(I)}_2+{\omega}^{(S)}_3$ are
satisfied. We then plot the corresponding values of the interfacial
wave number $\boldsymbol{k}^I_2$.  Schematically, this construction
process is depicted in figure (\ref{resonanceSchematic}).
\\ \indent The intercept (a) in figure (\ref{res}) corresponds to the
case when the interfacial wave copropagates in the direction of
surface waves; this precisely corresponds to the class III of
resonances studied in \cite{Alam}. We note that in this case the
interfacial waves excited are much longer and slower than the surface
wave, because comparatively it takes much less energy to distort the
interface between the layers than it does to lift and disturb the
upper layer. In the interfacial wave case the heavier fluid is lifted
in slightly less dense fluid, while for the surface waves the water is
lifted into the air.

\begin{figure}
  \begin{flushleft}
    \includegraphics[width=2.25in]{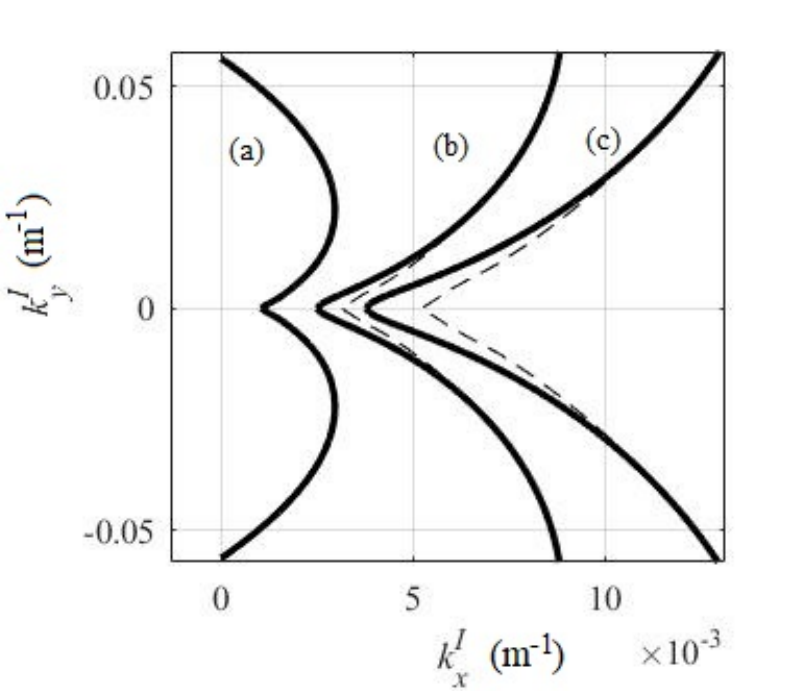}
    \caption{Resonances are not scale invariant: the solid curve
      depicts interfacial wavenumbers in resonance when one surface
      wave number is fixed to be (a) $\kappa=\frac{2\pi}{25}$
      $\text{m}^{-1}$, (b) $\kappa=\frac{3\times 2\pi}{25}$
      $\text{m}^{-1}$, (c) $\kappa=\frac{5\times 2\pi}{25}$
      $\text{m}^{-1}$. The dotted curves correspond to scaling curve
      (a) by factors of 3 and 5.}
    \label{ScaleInvar}
	\end{flushleft}
\end{figure}

Similarly, the intercept (b) in figure (\ref{res}) corresponds to
counter-propagating waves in class I resonance.  Notably, the
resonance curves are not scale invariant; figure (\ref{ScaleInvar})
depicts the resonance curve for the cases of fixed surface wave number
$\kappa =\frac{2j\pi}{25}$ $\text{m}^{-1}$, $j=1,3,5$, with the dotted
lines depicting the first resonance curve scaled by the respective
factors of three and five.  While the resonant manifold is
approximately scale invariant for large wavenumbers, scale invariance
is particularly violated near the class III collinear resonance. This
change in structure results in a different regime of dynamics for
longer surface wavelengths, as we will see in section
(\ref{SurfacePlaneWave}),

\section{Energy transfer in the JONSWAP spectrum}\label{EnergyTransferJONSWAPP}
\subsection{Surface plane wave}\label{SurfacePlaneWave}
\subsubsection{Formulation of the problem}
\indent Let us now consider a model problem of a single plane wave on
the surface of the top layer and an interfacial layer that is at rest
initially at $t=0$. While this problem may seem to be oversimplified,
it is motivated by real oceanographic scenarios. Indeed, when wind
blows on top of the ocean, the spectral energy density of the
generated surface gravity waves have a prominent narrow peak for a
specific wave number. We denote this wavevector of the initial surface
wave distribution by $\boldsymbol{\kappa}$, and assume the lower
interface is undisturbed except for small amplitude noise
$\epsilon_{n}$. Such a choice corresponds to the initial conditions
\begin{align}
\label{IC}
&n^{(S)}({\boldsymbol{k}},t=0)=\tilde{A}\:\delta({\boldsymbol{k}}-{\boldsymbol{\kappa}}), 
\\
&n^{(I)}({\boldsymbol{k}}, t=0)=O(\epsilon_{n}). 
\end{align}
In physical space such choice corresponds to 
\begin{align*}
\zeta_1({\boldsymbol{x}},t=0)=A \cos({\boldsymbol{\kappa}}\cdot {\boldsymbol{x}}-\omega_{\kappa} t),\:\: \zeta_1({\boldsymbol{x}},t=0)= \epsilon_{n}.
\end{align*}
In the calculations below, we take the surface wavenumber
$\boldsymbol{\kappa}$ to correspond to the peak wavenumber of the
JONSWAP spectrum \cite{JONSWAP}. Here, the 1-D JONSWAP spectrum can
be expressed in terms of $U_{19.5}$, the wind speed 19.5 m above the
ocean surface, and is given by
\begin{align}
        S(\omega)= \frac{\alpha g^2 } {\omega^5}\text{exp}\Big(-\frac{5}{4}\Big(\frac{\omega_0}{\omega}\Big)^4\Big)\gamma^r\ , \nonumber\\
r=\text{exp} (-\frac{(\omega-\omega_0)^2}{2\sigma^2 \omega_0^2}), \omega_0= g/U_{19.5}, \sigma=
    \begin{cases}
      0.07, & \text{if}\: \omega\leq\omega_0\\
      0.09, & \text{if}\: \omega > \omega_0
    \end{cases}.
        \label{1DJonswapp}
\end{align}

Consequently, the peak wavenumber $\boldsymbol{\kappa}$ and wave
amplitude, $A$, we use are determined solely by the speed of the wind.
Due to the surface spectrum being unidirectional, the dominant energy
exchange from the surface to interfacial spectrum occurs on and near
the class III resonance \cite{Alam}.

\indent Substituting \eqref{IC} into \eqref{gamma}, dropping terms
of order $\epsilon_n$, and keeping only the resonant and near resonant
terms, we obtain the following algebraic equations for the Boltzmann
rates of the interfacial and surface wave fields at $t=0$,
\begin{align}
&\gamma^{(I)}_{\boldsymbol{k}}(t=0) =-2\pi
\tilde{A}|J^{(SIS)}(\boldsymbol{\kappa},{\boldsymbol{k}},\boldsymbol{\kappa}-{\boldsymbol{k}})|^2
\mathcal{L}\big(\tilde{\omega}^{(S)}(\boldsymbol{\kappa})-\tilde{\omega}^{(I)}({\boldsymbol{k}})-\tilde{\omega}^{(S)}(\boldsymbol{\kappa}-{\boldsymbol{k}})\big),
\nonumber \\ \nonumber \\
&\gamma^{(S)}_{\boldsymbol{k}}(t=0)
=2\pi \tilde{A}|J^{(SIS)}(\boldsymbol{\kappa},\boldsymbol{\kappa}-{\boldsymbol{k}}, \boldsymbol{k})|^2 \mathcal{L}\big(\tilde{\omega}^{(S)}(\boldsymbol{\kappa})-\tilde{\omega}^{(I)}(\boldsymbol{\kappa}-{\boldsymbol{k}})-\tilde{\omega}^{(S)}({\boldsymbol{k}})\big),\nonumber \\
&\hskip 10
cm\boldsymbol{k}\in R_{III}.
\label{gammaSC}
\end{align}

\subsubsection{Unidirectional wave propagation}
\begin{figure}
  \centerline{\includegraphics[width=4in]{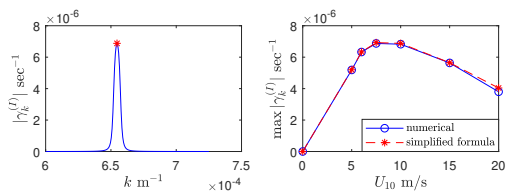}}
  \caption{(left) comparison of numerical solution of \eqref{gammaSC} and the analytical value of the peak growth rate using \eqref{gammaSimp} for a surface spectrum with wind speed of 10 m/s, (right) peak growth rate of interfacial waves vs. the
    wind speed for various values of windspeed according to both formulas.}
\label{freqfig}
\end{figure}

\indent We now consider the unidirectional wave propagation.  We find
the self-consistent value for $\gamma^{(I)}_{\boldsymbol{k}}$ by
numerically solving the $1D$ version of the algebraic equation
\eqref{gammaSC}. The numerical solution for the growth rate of
interfacial waves collinear to the surface wave is shown in figure
(\ref{freqfig}a) for a surface wavenumber of $\boldsymbol{\kappa}=2
\pi/19$ meters$^{-1}$.  Here the results were obtained by iterating
\eqref{gammaSC}.  The value of $\gamma^{(I)}_{\boldsymbol{k}}$ is
appears to be narrowly peaked around the resonant frequency
$\omega^{(I)}_{\boldsymbol{k}}=\omega^{(S)}_{\boldsymbol{\kappa}}-\omega^{(S)}_{\boldsymbol{\kappa}-\boldsymbol{k}}$.
We now vary the value of the wind speed and plot the magnitude of the
value of the peak of $\gamma^{(I)}({\bf k})$ as a function of the wind
speed. Results are shown on figure (\ref{freqfig}b). We see that for a
wind speed of roughly $8$ m/s there is a much more effective transfer
of energy to the interfacial waves, than at lower or higher wind
speeds. Interestingly, this wind speed is around the wind speed at
which whitecapping starts to occur.  Here the parameters used are
$\rho_u=1027 \text{ kg}/\text{m}^3$, $\Delta \rho=1 \text{ kg}/\text{
  m}^3$, $h_u=800$ m and $h_l=4000$ m. We can actually analytically
estimate the amplitude of the peak by the following arguments: If we
choose the interfacial wave number $\boldsymbol{k}_0$ so that the
resonance condition is satisfied,
i.e. $\omega^{(S)}_{\boldsymbol{\kappa}}-\omega^{(S)}_{\boldsymbol{\kappa}-\boldsymbol{k_0}}-\omega^{(I)}_{\boldsymbol{k}_0}=0$,
and assume that
$\gamma^{(S)}_{\boldsymbol{\kappa}},\gamma^{(S)}_{\boldsymbol{\kappa}-\boldsymbol{k}}\ll\gamma^{(I)}_{\boldsymbol{k}}$,
we obtain the following estimate for the growth rate of the resonant
interfacial wavenumber $\boldsymbol{k}_0$:
\begin{align}
\gamma^{(I)}_{\boldsymbol{k}_0}(t=0) =\sqrt{2\pi
\tilde{A}}|J^{(SIS)}(\boldsymbol{\kappa},{\boldsymbol{k}_0},\boldsymbol{\kappa}-{\boldsymbol{k}_0})|.
\label{gammaSimp}
\end{align}
The amplitude of gamma found by this equation are shown as red dots in
figures (\ref{freqfig} a),(\ref{freqfig}b); the result agrees with the
peak values given by iterating \eqref{gammaSC}, plotted in blue.

\begin{figure}
	\begin{flushleft}
		\includegraphics[width=4in]{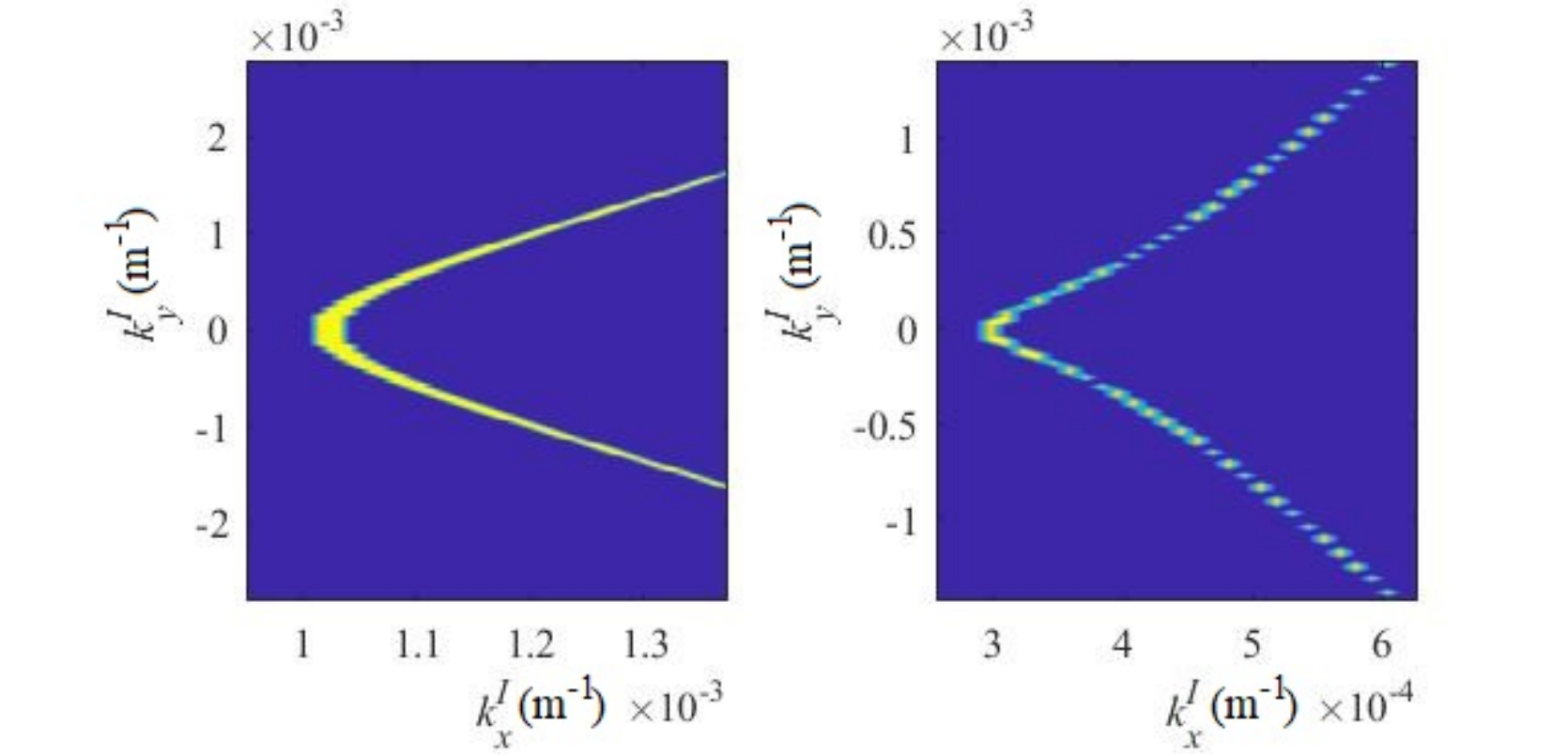}
 		\caption{(left) resonance curve as plotted in figure (\ref{ScaleInvar}) including physical width for $\lambda=20$ m surface spectrum, (right) $\lambda=80$ m surface spectrum.}
    \label{ResonanceWidth}
	\end{flushleft}
\end{figure}

Now that we have numerically evaluated the resonance width function
$\Gamma$, we can visualize the broadening of the resonant manifold.
So we broaden the resonant manifold (\ref{resfreq}) by allowing not
only resonant, but also near resonant interactions.  We therefore
replace resonant condition (\ref{resfreq}) by a more general condition
\begin{align}
  R_{III}= \{{|\boldsymbol{k}}_I:\tilde{\omega}^{(S)}({\boldsymbol{\kappa}})-\tilde{\omega}^{(I)}({\boldsymbol{k}}_I)-\tilde{\omega}^{(S)}({\boldsymbol{\kappa}}-{\boldsymbol{k}}_I))|<
  \Gamma^{(SIS)}_{\boldsymbol{\kappa},\boldsymbol{k}_I ,\boldsymbol{\kappa}-\boldsymbol{k}_I}.
  \label{generalResCond}
\end{align} 
\\
The results are depicted in in figure
\eqref{ResonanceWidth}, for the cases of the surface being a plane
wave of wavelength 20 meters and 80 meters. This figure replaces the
inset $(a)$ on figure (\ref{res}), and the figure (\ref{ScaleInvar}).
Here the amplitude of the plane wave is determined from the JONSWAP
spectrum.  We observe  that the biggest broadening of the
resonant curves occur at and  around the class
III collinear resonance.

\subsubsection{Unidirectional surface wave can generate oblique
  interfacial waves}
\indent We now remove the constraint of surface and interfacial
waves being collinear and consider the more general case of arbitrary
angle between them. Indeed, we observe that \eqref{gammaSimp} can be
evaluated for resonant 2-D interfacial wavenumbers which are not
necessarily collinear to $\boldsymbol{\kappa}$. Consequently, we
calculate the peak growth rate for interfacial waves all along the
resonance curve, of which the general shape is depicted in figure
(\ref{res}).
  
\begin{figure}
  \centerline{\includegraphics[width=3.5in]{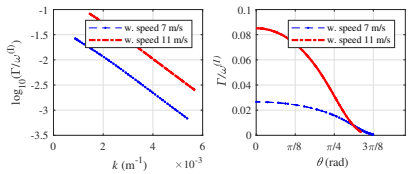}}
  \caption{The ratio between the resonance width
    $\Gamma^{(2)}_{\boldsymbol{\kappa},
      \boldsymbol{\kappa}-\boldsymbol{k}_I, \boldsymbol{k}_I}$ and the
    frequency of the excited interfacial wave
    ${\omega^{(I)}_{\boldsymbol{k}^I}}$ for two different wind speeds,
    as a function of wave number and angle along the resonance curve.}
  \label{reswidth}
  
\end{figure}

\begin{figure}
  \centerline{\includegraphics[width=1.75in]{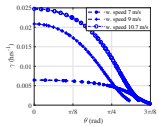}}
  \caption{The excitation rate, $\gamma_{\boldsymbol{k}}^{(I})$,
    plotted as a function of $\theta$, the angle between interfacial
    wave $\boldsymbol{k}$ and surface wave $ \boldsymbol{\kappa}$.}
  \label{gamwindangle}
\end{figure}

\indent In figures (\ref{reswidth}a),(\ref{reswidth}b) we plot the
ratio of the resonance width and the frequency of the interfacial wave
which is excited,
$\Gamma^{(SIS)}_{\boldsymbol{\kappa},
\boldsymbol{k}^I,\boldsymbol{\kappa}-\boldsymbol{k}_I}/{\omega^{(I)}_{\boldsymbol{k}^I}}$
along the entire resonance curve for the wind speeds 7 m/s and 10.7
m/s. We observe that for both wind speeds the low collinear wave
numbers experience the most resonance broadening, and that the
resonance width decreases roughly exponentially as a function
wavenumber.  Similarly, in figure (\ref{reswidth}c), (\ref{reswidth}d)
we plot the ratio of resonance width and frequency as a function of
the angle between the excited interfacial wave number and the fixed
surface wave number, verifying that it is largest when the angle is
small, though nonzero for other angles.
\\
\indent In figure (\ref{gamwindangle}) we plot the Boltzmann growth
rate of interfacial wave numbers along the 2-D resonance curve as a
function of the angle between the interfacial wave and the fixed
surface wave.  We see that a band of angles is excited on a timescale
similar to that of the peak collinear wave. Furthermore, as the wind
speed increases, the band of excited angles gets broader, so in
addition to the colinear waves in the direction of the wind speed, the
oblique waves are generated as well, making the mechanism of transfer
of energy from the wind to the internal waves much more effective.
\subsection{Analysis of matrix element}
\begin{figure}
  \centerline{\includegraphics[width=6in]{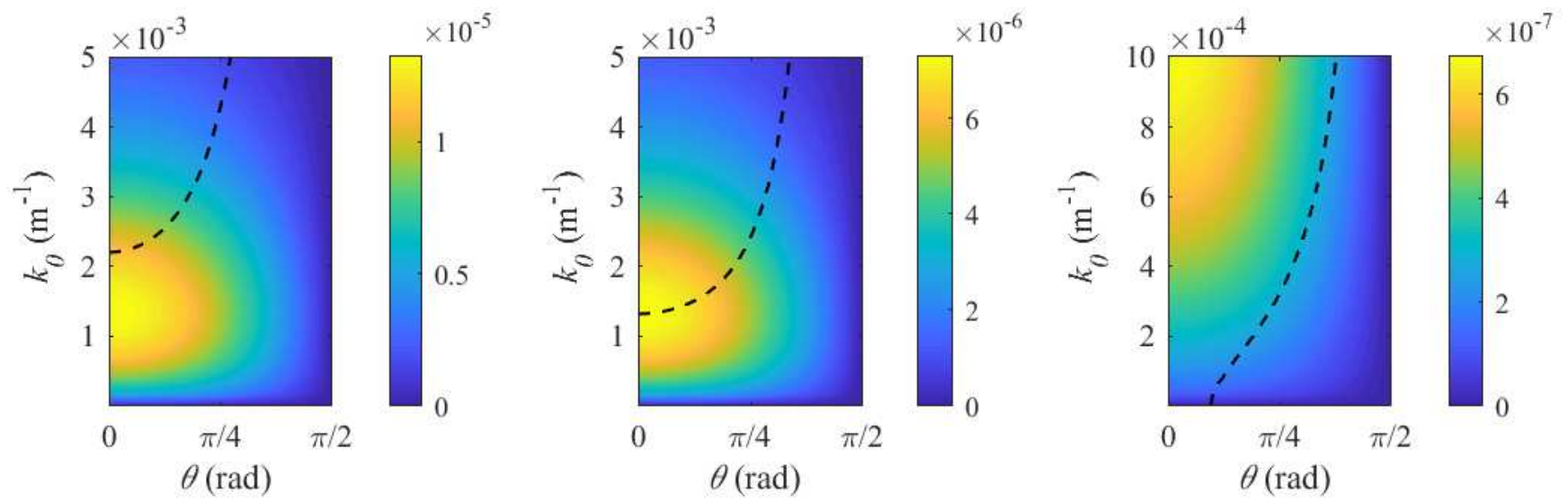}}
  \caption{The matrix element
    $J^{(SSI)}(\boldsymbol{\kappa},\boldsymbol{\kappa}+\boldsymbol{k}_I,\boldsymbol{k}_I)$
    with the resonant values of $k_I$ overlaid on top.  Surface
    wavelength is $\lambda=8$ m (left), $\lambda=14.5$ m (middle),
    $\lambda=150$ m (right).  We see a transition into a regime in
    which collinear resonances are no longer dominant, i.e. as the
    surface waves grow in wavelength, the resonant curve crosses a
    peak spatial frequency in which energy transfer is most
    efficient.
  }
\label{JSISFig}
\end{figure}

\indent The principle new feature of this paper is that we provide
first principle derivation of the magnitude of the strength of
interactions between the surface and interfacial waves. The value of
such interaction is called the matrix element, or interaction crossection
in wave turbulence theory. In figure (\ref{JSISFig}) we plot the
interaction coefficient $J^{(SIS)}$ which governs interaction strength
along resonances of the form depicted in figure (\ref{res}) for the
cases of surface wavelengths $\lambda = 8, 14.5, 150$ meters. Here we
fix the second surface wave number so that the resonance condition on
wave number \eqref{resk} is satisfied, plotting
$J^{(SIS)}(\boldsymbol{\kappa}+\boldsymbol{k}_I,\boldsymbol{k}_I,\boldsymbol{\kappa})$
as a function of the free wave number $\boldsymbol{k}_I$. The overlaid
white curve shows the resonant values of $\boldsymbol{k}_I$ determined
by the resonance condition on frequency. Combined, the contour plot
shows the interaction strength, with the resonance curve showing where
interactions are restricted to occur.  Notably, the interaction
coefficient is approximately scale invariant, as the structure remains
nearly the same for the case of surface waves with wavelength 9 meters
and surface swell waves with wavelength 150 meters.  In contrast, the
shape of the resonance curve experiences changes dependent on the
magnitude of the surface wave numbers, meaning that varying regimes
occur depending on where the resonance curve lies with respect to the
interaction coefficient.  For surface wavelength of approximately
$\lambda=14.5$ m the resonance curve aligns with the maximum value of
the matrix element, as seen in figure (\ref{JSISFig}b).  In this
regime, interactions as a whole seem most efficient.  Further beyond
this range the collinear resonance becomes less dominant, and for long
surface wavelengths, as in the case when $\lambda=150$ m, the maximum
growth rate is along noncollinear interfacial waves, as seen in figure
(\ref{JSISFig}c). 

\subsection{Excitation of interfacial waves by the JONSWAP surface wave spectrum, kinetic approach}
\subsubsection{Collinear Waves}
\begin{figure}
  \centerline{\includegraphics[width=3.5in]{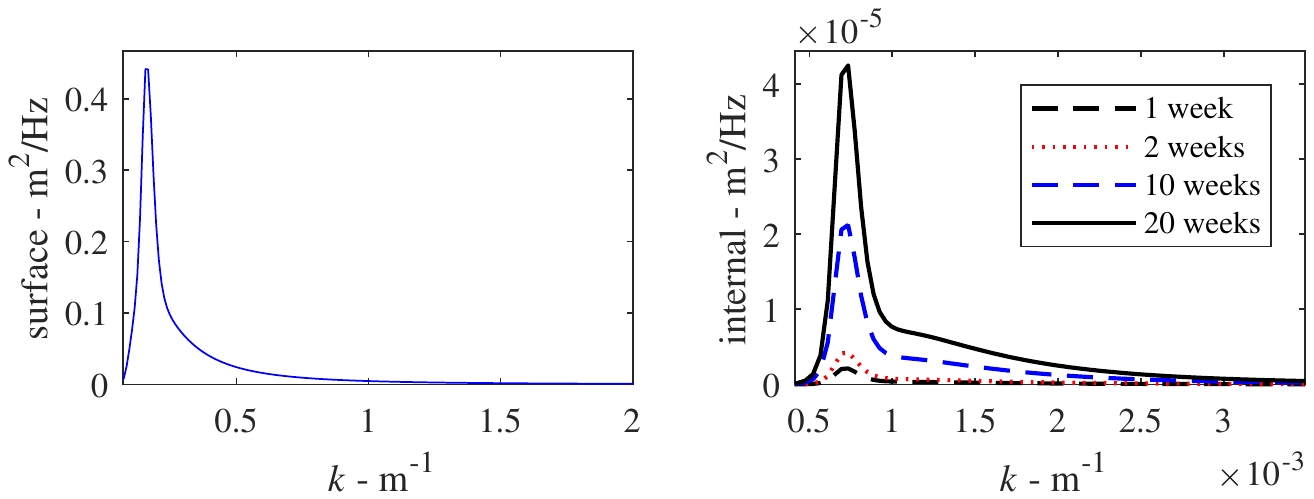}}
  \caption{{{(left) The fixed surface wave JONSWAP spectrum. (right) The interfacial wave spectrum over the course of twenty weeks, where $n^{(I)}_{\boldsymbol{k}}(t=0)=0$ initially.}}}
\label{specEv}
\end{figure}

The main motivation for this paper is to predict the transfer of
energy from the wind generated surface waves to the depth of the
ocean. Here we fix the spectrum of the surface waves to be JONSWAP
spectrum in the $x$ direction and homogeneous in the $y$ direction for
the wind speed of 15 m/s.  We assume that initially there are no
interfacial waves, i.e. $n^{(I)}(\boldsymbol{k},t=0)=0$.  To calculate
the growth rates, we iterate formula (\ref{gamma}) until a self
consistent value is obtained. We perform iterations until an iterate
is within $10^{-8}$ of the previous iterate.  After obtaining self
consistent values for the growth rates of each wavenumber, we then
evolve in time equations \eqref{KE} and plot the resulting spectrum of
the interfacial waves as a function of time in figure
(\ref{specEv}). Here the interfacial waves under wavelengths of $10$
meters are damped, the typical $10$ meter cut off for the internal
wave breaking. We see that the surface wave spectrum excites
interfacial waves on a time scale of days.  Furthermore, the relative
growth of the interfacial wave spectrum slows down over a period of
twenty weeks, with no visual difference between week nineteen and week
twenty other than near the peak frequency.

It appears that the spectral energy density of interfacial waves is a
``resoannt reflection'' of the spectra of the surface waves.  Indeed,
the spectrum on figure (\ref{specEv}) is defined solely by class III resonances
with the surface waves, and contributions from the interfacial wave
interactions is subdominant. It therefore appears that the
interactions between the interfacial waves is not an effective
mechanism for the redistribution of energy in the interfacial waves,
at least in the parameters chosen here. The spectral energy transfers
in the field of internal waves have been studied extensively
\cite{Muller86,ultraviolet17,regional,iwthPTN,LY,LvovTabak,LPT,LT}.
It is now understood that spectral energy transfer in internal waves
is dominated by the special class of nonlocal wave wave interactions,
called Induced Diffusion. These mechanism is absent in our model,
since it is a two layer system.

\subsection{Simulation of growth rates for continuous 2-D surface spectrum}
\begin{figure}
  \centerline{\includegraphics[width=5in]{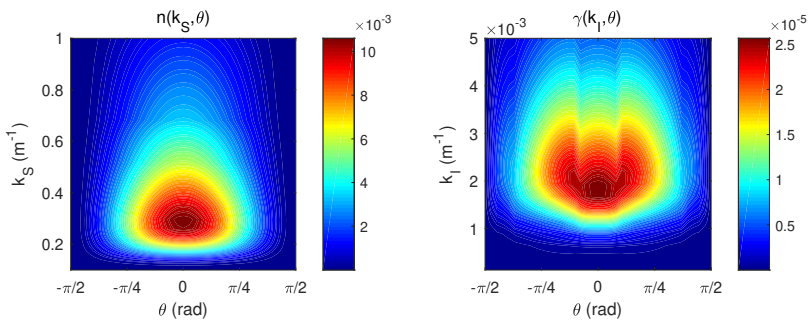}}
  \caption{(left) fixed surface spectrum (right) simulated growth rate of interfacial wave spectrum for wind speed 5 m/s}
  \label{gamma2d5}
\end{figure}

For the case when the surface spectrum is a more general 2-D spectrum,
we resort to numerically iterating \eqref{gamma}.  Here we consider the 2D version of equation \eqref{1DJonswapp} the JONSWAP
spectrum. Introducing simple angular dependence via a directional spreading function as described in \cite{janssenbook04}, we consider
\begin{equation}
  n^{(S)}(k,\theta) = A \cos^2 \theta\times  S(\omega^{(S)}_{\boldsymbol{k}})\frac{\text{d}\omega_{\boldsymbol{k}}}{\text{d}\boldsymbol{k}}, \:\: -\frac{\pi}{2}\leq \theta \leq \frac{\pi}{2},
  \label{Angular}
\end{equation}
where the spectrum is renormalized so that the total energy is the same as in the 1-D case, i.e. $A = \frac{\iint \cos^2 \theta\: S(\omega) \text{d}\omega \text{d}\theta}{\int S(\omega) \text{d}\omega}$.
We now substitute (\ref{Angular}) into (\ref{gamma}) and find the
self-consistent solution for $\gamma^{(I)}(k,\omega)$; the results
are shown in figure (\ref{gamma2d5}).

\begin{figure}
  \centerline{\includegraphics[width=5in]{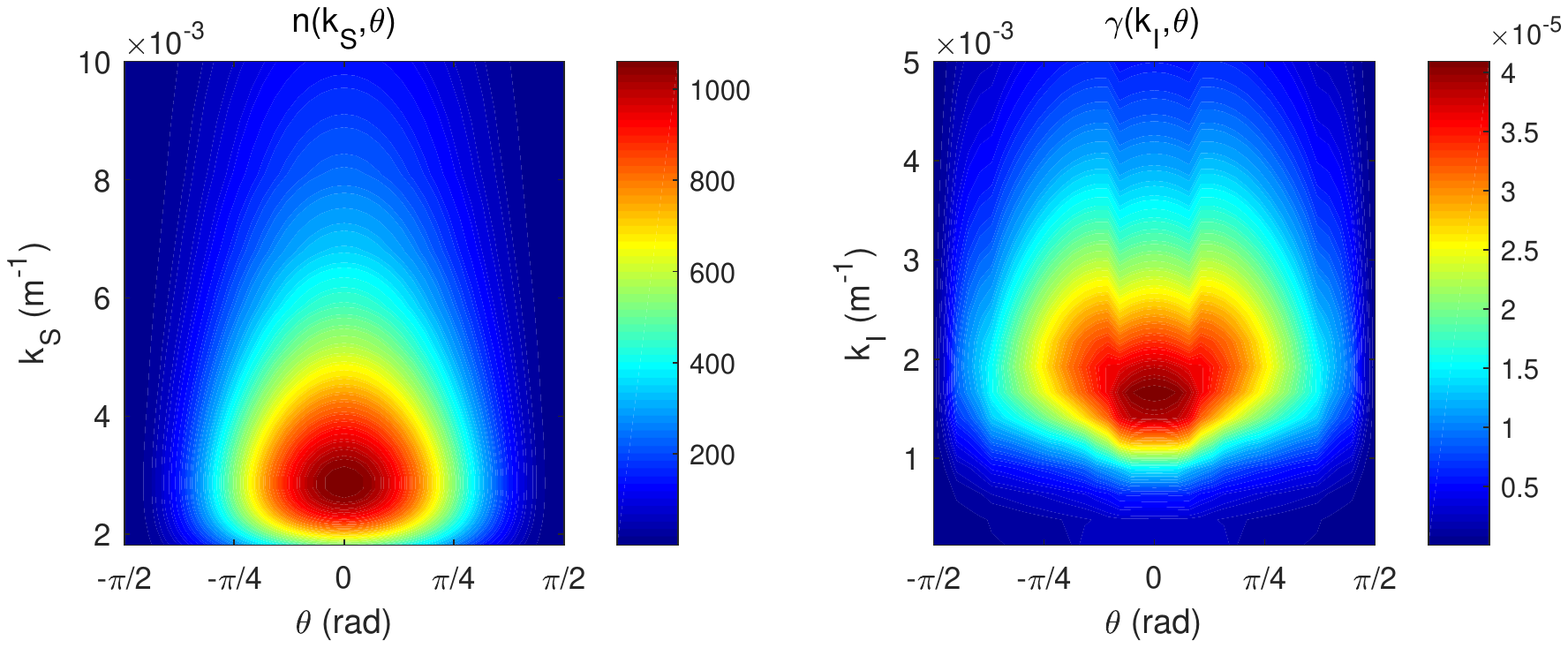}}
  \caption{(left) fixed surface spectrum (right) simulated growth rate of interfacial wave spectrum for hurricane wind speed 50 m/s}
    \label{gamma2d50}
  \end{figure}

We obtain self consistent values for the growth rate of interfacial
waves excited for the case of a 2-D P-M spectrum corresponding to a
wind speed of $5$ m/s in figure (\ref{gamma2d5}), and $50$ m/s in
figure (\ref{gamma2d50}).  Note that while these two pictures may look
similar, the scales of the figures are different. Higher wind speed
generates much larger band width of excited wave numbers, and the
growth rates are higher. Notably, the structure of the growth rate of
interfacial waves does not match the structure of the surface
spectrum.  There is a peak interfacial wave analogous to the peak of
the surface spectrum, but there are three lobes corresponding to
interactions where long surface waves are in resonance with oblique
interfacial waves.

Having calculated matrix element $J^{(SIS)}_{123}$, and the growth
rates of internal waves as a function of their wave number $k$ and
direction $\theta$, we are now in a position to simulate the full
evolving 2-D interfacial wave spectrum governed by \eqref{KE}. This is
the subject of future work.

\section{Discussion}\label{Discuss}

In the present paper we revisit the problem of the coupling of surface
gravity waves and an interfacial waves in a two layer system. We use
the recently developed Hamiltonian structure of the system
\cite{Choi2019} and systematically develop wave turbulence theory that
describes the time evolution of the spectral energy density of
waves. To achieve this goal, we first need to diagonalize quadratic
part of the Hamiltonian.  It appears to be a nontrivial task, which we
perform by a series of two canonical transformations. We end up with a
highly nontrivial over determined system of equations, which we solve,
thus finding normal modes of the system. We therefore set up a stage
for developing wave turbulence kinetic equations that describe
nonlinear spectral energy transfers between surface and interfacial
waves.

We then derive wave turbulence kinetic equations for the time
evolution of the spectral energy densities of such a two layer system.
Our kinetic equation allows not only resonant, but also near resonant
interactions. We revisit the question of possible three wave
resonances, and confirm that for normal sea conditions the resonant
picture developed in \cite{Ball, Thorpe, Alam} holds.  Interestingly,
the resonance condition still holds even for hurricane wind speeds and
the case of long surface swell; for long surface wavelengths, the
characteristic wavelength of the collinear interfacial waves becomes
much longer. This leads to resonant noncollinear interactions being
dominant, as evidenced by numerically studying the coupling
coefficient.

The Kinetic equation allows us to estimate the time scales of
excitation of interfacial waves by a single plain surface wave. We
find that under the assumption of the magnitude of the interfacial
waves being small, the growth rate is on the order of hours.  Notably,
we consider resonant interfacial waves in all directions, seeing that
the growth rate is maximized for the class III collinear resonance and
decreases with growing angle.  By using surface frequencies and
amplitudes estimated by the JONSWAP spectrum, we see a nonlinear
threshold of roughly 9 meters/second where energy transfer becomes
more effective, perhaps corresponding to conditions where the surface
waves transition into white capping.

We also consider the case when the spectrum of surface waves is given
by the continuous set of frequencies described in the JONSWAP
spectrum. We numerically solve the system of kinetic equations, and
find that interfacial waves are generated at a characteristic time
scale of days, and they eventually will reach a steady state at a
characteristic time of months.

Our future work will be focused on investigating in more detail the
interactions of interfacial and surface waves for the case of strong
wind, without limiting ourselves to collinear vectors. We also will 
attempt to use observations of internal waves and surface waves for
quantitative comparison with results given by the kinetic equation.
Lastly, we will attempt to generalize our kinetic equation to three or
more layers of the fluid.

\section*{Acknowledgments}
The authors are grateful Dr. Kurt L Polzin, Professor Wooyoung Choi
and Professor Gregor Kovacic for multiple useful discussions.
JZ and YL acknowledge support
from NSF OCE grant 1635866. YL acknowledges support from ONR grant
N00014-17-1-2852.  \appendix
\section{Nonlocal operator \cite{Choi2019}}\label{gamJ}
Here we list the components of the nonlocal operator and interaction matrix
elements derived in \cite{Choi2019}. These operators are constructed by multiplying
the Fourier transform of the quantity by the kernels given
below, then calculating the inverse Fourier transform.  {{The Fourier
kernels of operators as a function of wavenumber $k$, derived by \cite{Choi2019}, are
\begin{align}
\nonumber 
&J_k=\frac{1}{\rho_u \tanh(h_u k)\tanh(h_l k)+\rho_l},
\\
\nonumber 
&\gamma_{11,k}=k J_k \big[(\rho_l/\rho_u)\tanh(h_u k)+\tanh(h_l k)\big], \gamma_{12,k}=\gamma_{21,k}=k J_k \sech(h_u k) \tanh(h_l k), 
\\
\nonumber 
&\gamma_{22,k}=k J_k\tanh(h_l k), \gamma_{30,k}=J_k, \gamma_{31,k}=\sech(h_u k) J_k,
\\
\nonumber 
&\gamma_{32,k}=J_k\tanh(h_u k) \tanh(h_l k), \gamma_{33,k}=J_k(1+\tanh(h_u k) \tanh(h_l k)).
\end{align}}}

\section{Matrix elements in Fourier space \cite{Choi2019}} \label{MatrixElementFS} 
The coupling coefficients of the quadratic and cubic Hamiltonians from  \cite{Choi2019} are given
by
{{
\begin{align*}
h^{(1a)}(\boldsymbol{k})&=\rho_u g,
\\
h^{(2a)}(\boldsymbol{k})&=\frac{(\rho_l/\rho_u)k\tanh(h_uk)+k \tanh(h_l k)}{\rho_u \tanh(h_u k)\tanh(h_l k)+\rho_l},
\\
h^{(3a)}(\boldsymbol{k})&=\Delta \rho g,
\\
h^{(4a)}(\boldsymbol{k})&=\frac{k\tanh(h_l k)}{\rho_u \tanh(h_u k)\tanh(h_l k)+\rho_l},
\\
h^{(5a)}(\boldsymbol{k}_1,\boldsymbol{k}_2)&=\frac{k_1\tanh(h_l k_1)\sech(h_u k_1)}{\rho_u \tanh(h_u k_1)\tanh(h_l k_1)+\rho_l}+\frac{k_2\tanh(h_l k_2)\sech(h_u k_2)}{\rho_u \tanh(h_u k_2)\tanh(h_l k_2)+\rho_l}, 
\end{align*}
\begin{align*}
&h^{(1)}_{123}=-\frac{1}{2}(\vec{k_1}\cdot\vec{k_2})/\rho_u
-\frac{1}{2} k_1 k_2 \frac{(\rho_l\tanh(h_u k_1)+\rho_u\tanh(h_l k_1))(\rho_l\tanh(h_u k_2)+\rho_u\tanh(h_l k_2))}{\rho_u(\rho_u \tanh(h_uk_1)\tanh(h_lk_1)+\rho_l)(\rho_u \tanh(h_uk_2)\tanh(h_lk_2)+\rho_l)},
\\
&h^{(2)}_{123}=-k_1 k_2 \frac{\sech(h_uk_2)\tanh(h_l k_2)(\rho_l\tanh(h_u k_1)+\rho_u \tanh(h_l k_1))}{(\rho_u \tanh(h_uk_1)\tanh(h_lk_1)+\rho_l)(\rho_u \tanh(h_uk_2)\tanh(h_lk_2)+\rho_l)},
\\
&h^{(3)}_{123}=-\frac{1}{2}\rho_u k_1 k_2 \frac{\sech(h_u k_1)\sech(h_uk_2)\tanh(h_l k_1)\tanh(h_l k_2)}{(\rho_u \tanh(h_uk_1)\tanh(h_lk_1)+\rho_l)(\rho_u \tanh(h_uk_2)\tanh(h_lk_2)+\rho_l)},
\\
&h^{(4)}_{123}=-\frac{1}{2}\Delta \rho \frac{\sech(h_u k_1)\sech(h_u k_2)}{(\rho_u \tanh(h_uk_1)\tanh(h_lk_1)+\rho_l)(\rho_u \tanh(h_uk_2)\tanh(h_lk_2)+\rho_l)}
\\
&\times\big[-(\rho_l/\rho_u)(\vec{k_1}\cdot\vec{k_2})+k_1 k_2 \tanh(h_l k_1)\tanh(h_l k_2)\big],
\\
&h^{(5)}_{123}=-\frac{\sech(h_u k_1)}{(\rho_u \tanh(h_uk_1)\tanh(h_lk_1)+\rho_l)(\rho_u \tanh(h_uk_2)\tanh(h_lk_2)+\rho_l)}
\\
&\times \big[\Delta \rho \: k_1 k_2 \tanh(h_l k_1)\tanh(h_l k_2)+\rho_l (1+\tanh(h_u k_2)\tanh(h_l k_2))(\vec{k_1}\cdot \vec{k_2})\big],
\\
&h^{(6)}_{123}=-\frac{1}{2(\rho_u \tanh(h_uk_1)\tanh(h_lk_1)+\rho_l)(\rho_u \tanh(h_uk_2)\tanh(h_lk_2)+\rho_l)}\\
&\times 
\big[\Delta \rho \: k_1 k_2 \tanh(h_l k_1)\tanh(h_l k_2)
\\
&+(\rho_l - \rho_u \tanh(h_u k_1)\tanh(h_u k_2)\tanh(h_l k_1)\tanh(h_l k_2))(\vec{k_1}\cdot \vec{k_2})\big]. 
\end{align*}
}}
\section{Canonical transformation}\label{ME}
\subsection{Linear transformation}
The linearized equations of motion for
\begin{align*}
\begin{pmatrix}\boldsymbol{u}\\
\boldsymbol{v}\end{pmatrix}
=
    \begin{pmatrix}
    \zeta^{(1)}\\
    \zeta^{(2)}\\
    \Psi^{(1)}\\
    \Psi^{(2)}
    \end{pmatrix} 
\end{align*}
can be written in the form
\begin{align*}
    \begin{pmatrix}\dot{\boldsymbol{u}}\\
\dot{\boldsymbol{v}}\end{pmatrix}=\begin{pmatrix} O_{2\times2}& G_{2\times2}\\
    M_{2\times2}& O_{2\times 2}
    \end{pmatrix} 
    \begin{pmatrix}\boldsymbol{u}\\
\boldsymbol{v}\end{pmatrix}.
\end{align*}
    We seek a canonical transformation to normal modes of the
    system.  The condition of the linear transformation to be
    canonical is that it is representable as
\begin{align*}
    \begin{pmatrix}\dot{\boldsymbol{u}}\\
\dot{\boldsymbol{v}}\end{pmatrix}=\begin{pmatrix} O_{2\times2}& I_{2\times2}\\
    -I_{2\times2}& O_{2\times2}
    \end{pmatrix}
    \begin{pmatrix}\boldsymbol{u}\\
\boldsymbol{v}\end{pmatrix}, 
\end{align*} 
with $G$ and $M$ such that each are diagonalizable via unitary
similarity transformations.  However, for the system under
consideration this is not possible, and we substitute a more general
linear transformation in the complex action variables of the form
\cite{ZLF}
\begin{subequations}
\begin{align}
&a^{(U)}_k=Q^{(1)}_{\boldsymbol{k}} c^{(I)}_{\boldsymbol{k}}+Q^{(2)}_{\boldsymbol{k}} c^{(I)*}_{\boldsymbol{-k}}+Q^{(3)}_{\boldsymbol{k}} c^{(S)}_{\boldsymbol{k}}+Q^{(4)}_k c^{(S)*}_{\boldsymbol{-k}},
\\
&a^{(L)}_k=Q^{(5)}_{\boldsymbol{k}} c^{(I)}_{\boldsymbol{k}}+Q^{(6)}_{\boldsymbol{k}} c^{(I)*}_{\boldsymbol{-k}}+Q^{(7)}_{\boldsymbol{k}} c^{(S)}_{\boldsymbol{k}}+Q^{(8)}_k c^{(S)*}_{\boldsymbol{-k}}. 
\label{transformation}
\end{align}
\end{subequations}
In order to preserve the Hamiltonian structure during the
transformation of the Hamiltonian, the transformation should be
canonical. Conditions for transformations to be canonical are given by
\cite{ZLF}:
\begin{subequations} 
\begin{align}
\label{condfirst}
&|Q^{(1)}_{\boldsymbol{k}}|^2-|Q^{(2)}_{\boldsymbol{k}}|^2+|Q^{(3)}_{\boldsymbol{k}}|^2-|Q^{(4)}_{\boldsymbol{k}}|^2=1,
\:\:
|Q^{(5)}_{\boldsymbol{k}}|^2-|Q^{(6)}_{\boldsymbol{k}}|^2+|Q^{(7)}_{\boldsymbol{k}}|^2-|Q^{(8)}_{\boldsymbol{k}}|^2=1,
\\
&Q^{(1)}_{\boldsymbol{k}}Q^{(5)*}_{\boldsymbol{k}}+Q^{(3)}_{\boldsymbol{k}}Q^{(7)*}_{{\boldsymbol{k}}}=Q^{(2)}_{\boldsymbol{k}}Q^{(6)*}_{\boldsymbol{k}}+Q^{(4)}_{\boldsymbol{k}}Q^{(8)*}_{\boldsymbol{k}},
\:\:
Q^{(1)}_{\boldsymbol{k}}Q^{(6)}_{-{\boldsymbol{k}}}+Q^{(3)}_{\boldsymbol{k}}Q^{(8)}_{-{\boldsymbol{k}}}=Q^{(2)}_kQ^{(5)}_{-{\boldsymbol{k}}}+Q^{(4)}_kQ^{(7)}_{-{\boldsymbol{k}}},
\\
&Q^{(1)}_{\boldsymbol{k}}Q^{(2)}_{-{\boldsymbol{k}}}+Q^{(3)}_{\boldsymbol{k}}Q^{(4)}_{-{\boldsymbol{k}}}=Q^{(1)}_{-{\boldsymbol{k}}}Q^{(2)}_{\boldsymbol{k}}+Q^{(3)}_{-{\boldsymbol{k}}}Q^{(4)}_{\boldsymbol{k}},
\:\:
Q^{(5)}_{\boldsymbol{k}}Q^{(6)}_{-{\boldsymbol{k}}}+Q^{(7)}_{\boldsymbol{k}}Q^{(8)}_{-{\boldsymbol{k}}}=Q^{(5)}_{-{\boldsymbol{k}}}Q^{(6)}_{\boldsymbol{k}}+Q^{(7)}_{-{\boldsymbol{k}}}Q^{(8)}_{\boldsymbol{k}}. 
\label{condtriv} 
\end{align}
\end{subequations}
In order to diagonalize the quadratic part of the system,
we further require that terms of the form
$c_{\boldsymbol{k}}^*C_{\boldsymbol{k}}$,
$c_{\boldsymbol{k}}C_{-{\boldsymbol{k}}}$,
$c_{\boldsymbol{k}} c_{-\boldsymbol{k}}$ and
$C_{\boldsymbol{k}} C_{-\boldsymbol{k}}$ and their complex conjugates
vanish, which lead to the the additional conditions

 \begin{subequations}
\begin{align}
&\omega^{(1)}_{\boldsymbol{k}}(Q^{(1)}_{\boldsymbol{k}}Q^{(3)*}_{\boldsymbol{k}}+Q^{(2)*}_{-{\boldsymbol{k}}}Q^{(4)}_{-{\boldsymbol{k}}})+\omega^{(2)}(Q^{(5)}_{\boldsymbol{k}}Q^{(7)*}_{\boldsymbol{k}}+Q^{(6)*}_{-{\boldsymbol{k}}}Q^{(8)}_{-{\boldsymbol{k}}})
\nonumber
\\
&+F_{\boldsymbol{k}}\big[(Q^{(1)}_{\boldsymbol{k}}-Q^{(2)*}_{-{\boldsymbol{k}}})(Q^{(8)}_{-{\boldsymbol{k}}}-Q^{(7)*}_{\boldsymbol{k}})
+(Q^{(3)*}_{\boldsymbol{k}}-Q^{(4)}_{-{\boldsymbol{k}}})(Q^{(6)*}_{-{\boldsymbol{k}}}-Q^{(5)}_{\boldsymbol{k}})\big]=0,
\label{condcancel1}
\\
\nonumber
\\
\nonumber
&\omega^{(1)}_{\boldsymbol{{\boldsymbol{k}}}}(Q^{(1)}_{\boldsymbol{k}} Q^{(4)*}_{\boldsymbol{k}}+Q^{(2)*}_{-{\boldsymbol{k}}}Q^{(3)}_{-{\boldsymbol{k}}})+\omega^{(2)}_{\boldsymbol{k}}(Q^{(5)}_{\boldsymbol{k}}Q^{(8)*}_{\boldsymbol{k}}+Q^{(6)*}_{-{\boldsymbol{k}}}Q^{(7)}_{-{\boldsymbol{k}}})
\\
&+F_{\boldsymbol{k}}\big[(Q^{(5)}_{\boldsymbol{k}}-Q^{(6)*}_{-{\boldsymbol{k}}})(Q^{(3)}_{-{\boldsymbol{k}}}-Q^{(4)*}_{\boldsymbol{k}})
+(Q^{(7)}_{-{\boldsymbol{k}}}-Q^{(8)*}_{\boldsymbol{k}})(Q^{(1)}_{\boldsymbol{k}}-Q^{(2)*}_{-{\boldsymbol{k}}})
\big]=0,
\\
\nonumber
\\
&\omega^{(1)}_{\boldsymbol{k}} Q^{(1)}_{\boldsymbol{k}} Q^{(2)*}_{\boldsymbol{k}}+
\omega^{(2)}_{\boldsymbol{k}} Q^{(5)}_{\boldsymbol{k}} Q^{(6)*}_{\boldsymbol{k}}
+F_{\boldsymbol{k}}\big[Q^{(1)}_{\boldsymbol{k}}(Q^{(5)}_{-{\boldsymbol{k}}}-Q^{(6)*}_{\boldsymbol{k}})+Q^{(2)*}_{\boldsymbol{k}}(Q^{(6)*}_{-{\boldsymbol{k}}}-Q^{(5)}_{\boldsymbol{k}})\big]=0,
\\
\nonumber
\\
&\omega^{(1)}_{\boldsymbol{k}} Q^{(3)}_{\boldsymbol{k}} Q^{(4)*}_{\boldsymbol{k}}+
\omega^{(2)}_{\boldsymbol{k}} Q^{(7)}_{\boldsymbol{k}} Q^{(8)*}_{\boldsymbol{k}}
+F_{\boldsymbol{k}}\big[Q^{(3)}_{\boldsymbol{k}}(Q^{(7)}_{-{\boldsymbol{k}}}-Q^{(8)*}_{\boldsymbol{k}})+Q^{(4)*}_{\boldsymbol{k}}(Q^{(8)*}_{-{\boldsymbol{k}}}-Q^{(7)}_{\boldsymbol{k}})\big]=0.
\label{condcancel2}
\end{align} 
\end{subequations} 
\indent We note that these conditions give a transformation of a form
analogous to that of the Bogoliubov---Valatin transformation widely
used to diagonalize Hamiltonians in quantum mechanics
\cite{Bog58,Valatin58}.

We solve for the coefficients of the transformation---the full system
\eqref{condfirst}---\eqref{condcancel2} to obtain the Hamiltonian of
the system in the diagonal form. This system contains eight complex
and two real nonlinear coupled equations for eight complex unknowns,
which makes this task nontrivial.  It might seem that the the system
is overdetermined; yet it turns out this is not the case.  Namely,
under the additional assumption that
  \begin{equation}
  Q^{(i)}_{\boldsymbol{k}},\:
      i=1,2...\:8 {\rm \: are\  real\  even\  functions\  of\ }\boldsymbol{k},
\label{Assumption}
\end{equation}
equations \eqref{condtriv} are trivially satisfied. We then are left
with six complex equations and two real equations for eight complex
unknowns.  We obtain a particular solution to
\eqref{condfirst}---\eqref{condcancel2},
\begin{align}
&a^{(U)}_{\boldsymbol{k}}= \sin \theta[(\cosh \phi)  c^{(I)}_{\boldsymbol{k}} + (\sinh \phi) c^{(I)*}_{-{\boldsymbol{k}}}]+\cos \theta [(\cosh \psi) c^{(S)}_{\boldsymbol{k}} + (\sinh \psi) c^{(S)*}_{-{\boldsymbol{k}}}],
\nonumber
\\
&a^{(L)}_{\boldsymbol{k}}=\cos \theta[(\alpha\cosh \phi   + \beta\sinh \phi)c^{(I)}_{\boldsymbol{k}}+(\alpha\sinh \phi + \beta \cosh \phi) c^{(I)*}_{-{\boldsymbol{k}}}]
\nonumber
\\
&- \sin \theta [(\alpha\cosh \psi   + \beta\sinh \psi)c^{(S)}_{\boldsymbol{k}}+(\alpha\sinh \psi + \beta \cosh \psi) c^{(S)*}_{-{\boldsymbol{k}}}].
\label{finaltrans}
\end{align}
Here $\alpha, \beta, \theta, \phi,$ and $\psi$ are given by \eqref{transcoeff1}---\eqref{transcoeff2}.  Additional
detailed analyses shows that the general solution has an additional
{\it six} complex random phases, which are taken here to be
zero. Nonzero phases alters the phases of $c^{(S)}_{\boldsymbol{k}}$
and $c^{(I)}_{\boldsymbol{k}}$, but do not change the corresponding
linear dispersion relationships or the strength of coupling of the
normal modes.

We emphasize that the transformation is expressed in closed form in
terms of $\omega^{(1)}_{\boldsymbol{k}},
\omega^{(2)}_{\boldsymbol{k}},$ and $F_{\boldsymbol{k}}$, and that
this approach works for a general system with two types of interacting
waves and quadratic coupling term.

\subsection{Transformation coefficients}
Define the following functions: \begin{align*}
C^{(1)}_{\boldsymbol k}=-\frac{F^{(1)}_{\boldsymbol{k}}+F^{(2)}_{\boldsymbol{k}}}{2\sqrt{F^{(1)}_{\boldsymbol{k}}F^{(2)}_{\boldsymbol{k}}}},
C^{(2)}_{\boldsymbol k}=-\frac{F^{(1)}_{\boldsymbol{k}}-F^{(2)}_{\boldsymbol{k}}}{2\sqrt{F^{(1)}_{\boldsymbol{k}}F^{(2)}_{\boldsymbol{k}}}},
\theta_{\boldsymbol k}=\frac{1}{2}\arctan\Big[\frac{-4F^{(3)}_{\boldsymbol{k}} \sqrt{F^{(1)}_{\boldsymbol{k}}F^{(2)}_{\boldsymbol{k}}}}{{F^{(1)}_{\boldsymbol{k}}}^2-{F^{(2)}_{\boldsymbol{k}}}^2}\Big],
\end{align*}
\begin{align*}
    &\mu_{\boldsymbol k}=F^{(2)}_{\boldsymbol{k}} C^{(1)}_{\boldsymbol k} C^{(2)}_{\boldsymbol k}\cos^2\theta_{\boldsymbol k}+F^{(3)}_{\boldsymbol k}(C^{(1)}_{\boldsymbol k}-C^{(2)}_{\boldsymbol k})\sin\theta_{\boldsymbol k}\cos\theta_{\boldsymbol k},
    \\
    &\sigma_{\boldsymbol k}=F^{(2)}_{\boldsymbol{k}} C^{(1)}_{\boldsymbol k} C^{(2)}_{\boldsymbol k}\sin^2\theta_{\boldsymbol{k}}-F^{(3)}_{\boldsymbol{k}}(C^{(1)}_{\boldsymbol k}-C^{(2)}_{\boldsymbol k})\sin\theta_{\boldsymbol{k}}\cos\theta_{\boldsymbol{k}}.
\end{align*}
Then we obtain
\begin{align}
    &\alpha_{\boldsymbol{k}}=F^{(1)}_{\boldsymbol{k}}\sin^2\theta_{\boldsymbol{k}}+F^{(2)}_{\boldsymbol{k}}({C^{(1)}_{\boldsymbol k}}^2+{C^{(2)}_{\boldsymbol k}}^2)\cos^2\theta_{\boldsymbol{k}}-2F^{(3)}_{\boldsymbol k}({C^{(1)}_{\boldsymbol k}}-{C^{(2)}_{\boldsymbol k}})\sin\theta_{\boldsymbol{k}}\cos\theta_{\boldsymbol{k}},
    \\
    &\beta_{\boldsymbol{k}}=F^{(1)}_{\boldsymbol{k}}\cos^2\theta+F^{(2)}_{\boldsymbol{k}}({C^{(1)}_{\boldsymbol k}}^2+{C^{(2)}_{\boldsymbol k}}^2)\sin^2\theta_{\boldsymbol{k}}+2F^{(3)}_{\boldsymbol k}({C^{(1)}_{\boldsymbol k}}-{C^{(2)}_{\boldsymbol k}})\sin\theta_{\boldsymbol{k}}\cos\theta_{\boldsymbol{k}},
    \label{transcoeff1}
\end{align}
\begin{align}
\phi_{\boldsymbol{k}}=\frac{1}{2}\tanh^{-1}\big(-\frac{2\mu_{\boldsymbol{k}}}{\alpha_{\boldsymbol{k}}}\big),\psi_{\boldsymbol{k}}=\frac{1}{2}\tanh^{-1}\big(-\frac{2\sigma_{\boldsymbol{k}}}{\beta_{\boldsymbol{k}}}\big).
\label{transcoeff2}
\end{align}
Thus for the Hamiltonian to be diagonalizible we have the condition that 
\begin{align*}
    -1< -\frac{2\mu_{\boldsymbol{k}}}{\alpha_{\boldsymbol{k}}}<1,\:\: -1<-\frac{2\sigma_{\boldsymbol{k}}}{\beta_{\boldsymbol{k}}}<1.
\end{align*}

\section{Matrix elements for normal mode interactions}\label{METrans}
Define the following functions written in terms of the matrix elements from \cite{Choi2019}, 
\begin{equation*}
f^{(1)}_k=\sqrt{[h^{(1a)}_k]^{-1}h^{(2a)}_k}, f^{(2)}_k=\sqrt{[h^{(3a)}_k]^{-1}h^{(4a)}_k}.
\end{equation*}
Then the matrix elements {\it{before}} applying the canonical transformation \eqref{finaltrans} are
\begin{align*}
&G^{(1)}_{123}=-\frac{h^{(1)}_{123}}{2}\sqrt{\frac{f^{(1)}_3}{2f^{(1)}_1f^{(1)}_2}},
G^{(2)}_{123}=-\frac{h^{(2)}_{123}}{2}\sqrt{\frac{f^{(1)}_3}{2f^{(1)}_1f^{(2)}_2}},
\\
&G^{(3)}_{123}=-\frac{h^{(3)}_{123}}{2}\sqrt{\frac{f^{(1)}_3}{2f^{(2)}_1f^{(2)}_2}},
G^{(4)}_{123}=-\frac{h^{(4)}_{123}}{2}\sqrt{\frac{f^{(2)}_3}{2f^{(1)}_1f^{(1)}_2}},
\\
&G^{(5)}_{123}=-\frac{h^{(5)}_{123}}{2}\sqrt{\frac{f^{(2)}_3}{2f^{(1)}_1f^{(2)}_2}},
G^{(6)}_{123}=-\frac{h^{(6)}_{123}}{2}\sqrt{\frac{f^{(2)}_3}{2f^{(2)}_1f^{(2)}_2}},
\\
&V^{(1)}_k=G^{(1)}_{12-3}-G^{(1)}_{1-32}-G^{(1)}_{-321},
\\
&V^{(2)}_k=G^{(2)}_{12-3}-G^{(2)}_{1-32}-G^{(2)}_{-321}, \: ...
\\
&V^{(i)}_k=G^{(i)}_{12-3}-G^{(i)}_{1-32}-G^{(i)}_{-321}, \text{for\:}j=1,2,3...6.
\end{align*}

To calculate the matrix elements after applying transformation
\eqref{finaltrans} we make use of the following permutation operator to
shorten notation,
\begin{align*}
    P^{123}_{ijk}G=\sum_{i\neq j\neq k} G_{ijk},
\end{align*}
where the summation is taken for $i,j,k\in \{1,2,3\}$. Furthermore, to
shorten expressions involving products of the transformation
coefficients \eqref{transformation} we use the shorthand notation
$Q^{ijk}\equiv Q^{(i)}_1 Q^{(j)}_2 Q^{(k)}_3, \: i,j,k=1,2,3...8$.
\\ \indent We obtain the coupling coefficients
\begin{align}
    &J^{(S_1 I_2 S_3)}_{123}=
    \nonumber
    \\
    &Q^{342} P^{1-2-3}_{ijk}G^{(1)}
    +(Q^{386} P^{1-2-3}_{ij1}+Q^{746} P^{1-2-3}_{ij2}+Q^{782} P^{1-2-3}_{ij3}) G^{(3)}
    \nonumber
    \\
    &+(Q^{742} P^{1-2-3}_{ij1}+Q^{382} P^{1-2-3}_{ij2}+Q^{346} P^{1-2-3}_{ij3})G^{(4)}+Q^{786} P^{1-2-3}_{ijk} G^{(6)}
    \nonumber 
    \\
    &(Q^{442}P^{-1-2-3}_{ij1}+Q^{332} P^{12-3}_{ij2}+Q^{341} P^{1-23}_{ij3}) V^{(1)}
   +(Q^{486} P^{-1-2-3}_{ij1}
    \nonumber
    \\
    &+Q^{736} P^{12-3}_{ij2}+Q^{781} P^{1-23}_{ij3}) V^{(3)}
    +(Q^{842} P^{-1-2-3}_{ij1}+Q^{372} P^{12-3}_{ij2}
    \nonumber
    \\
    &+Q^{345} P^{1-23}_{ij3}) V^{(4)}
    +(Q^{886} P^{-1-2-3}_{ij1}+Q^{776} P^{12-3}_{ij2}+Q^{785}P^{1-23}_{1j3})V^{(6)}, 
    \label{meSIS}
\end{align}
\begin{align}
    &J^{(I_1 I_2 I_3)}_{123}=
    \nonumber
    \\
    &Q^{112} P^{1-2-3}_{ijk}G^{(1)}
    +(Q^{386} P^{1-2-3}_{ij1}+Q^{746} P^{1-2-3}_{ij2}+Q^{782} P^{1-2-3}_{ij3}) G^{(3)}
    \nonumber
    \\
    &+(Q^{742} P^{1-2-3}_{ij1}+Q^{382} P^{1-2-3}_{ij2}+Q^{346} P^{1-2-3}_{ij3})G^{(4)}+Q^{786} P^{1-2-3}_{ijk} G^{(6)}
    \nonumber 
    \\
    &(Q^{442}P^{-1-2-3}_{ij1}+Q^{332} P^{12-3}_{ij2}+Q^{341} P^{1-23}_{ij3}) V^{(1)}
   +(Q^{486} P^{-1-2-3}_{ij1}
    \nonumber
    \\
    &+Q^{736} P^{12-3}_{ij2}+Q^{781} P^{1-23}_{ij3}) V^{(3)}
    +(Q^{842} P^{-1-2-3}_{ij1}+Q^{372} P^{12-3}_{ij2}
    \nonumber
    \\
    &+Q^{345} P^{1-23}_{ij3}) V^{(4)}
    +(Q^{886} P^{-1-2-3}_{ij1}+Q^{776} P^{12-3}_{ij2}+Q^{785}P^{1-23}_{1j3})V^{(6)}. 
\end{align}

\bibliographystyle{jfm}

\end{document}